\documentclass[acmsmall,screen,nonacm]{acmart}
\usepackage{graphicx} 
\usepackage{amsmath,amsfonts} 

\usepackage{hyperref}  
\usepackage{url}       
\usepackage{flushend}  
\usepackage{pifont}      
\usepackage{graphicx}    
\usepackage{comment}
\usepackage{booktabs}    
\usepackage{array}       
\usepackage{multirow}
\usepackage{adjustbox} 
\usepackage{tcolorbox}
\usepackage{subcaption}
\usepackage[table]{xcolor}
\usepackage{enumitem}
\usepackage{xcolor}

\title{Test Case Selection for Deep Neural Networks: A Replication Study on LLMs for Code}

\author{Ali Asgari}
\email{a.asgari-2@tudelft.nl}
\affiliation{%
  \institution{Delft University of Technology}
  \country{The Netherlands}
}

\author{Mitchell Olsthoorn}
\email{m.j.g.olsthoorn@tudelft.nl}
\affiliation{%
  \institution{Delft University of Technology}
  \country{The Netherlands}
}

\author{Annibale Panichella}
\email{a.panichella@tudelft.nl}
\affiliation{%
  \institution{Delft University of Technology}
  \country{The Netherlands}
}

\renewcommand\footnotetextcopyrightpermission[1]{}

\begin{document}

\begin{abstract}
Recently, test case selection (TCS) techniques have been explored to support the operational evaluation of deep neural networks (DNNs) under limited testing budgets, where labeling cost is a primary concern and uncovering model failures early is a key objective. Although prior studies report promising results, existing empirical evaluations focus almost exclusively on vision-based DNNs and datasets. As observed in recent surveys, models and datasets specifically designed for software engineering tasks have not been considered, leaving it unclear whether prior findings generalize to LLM code models.

This paper presents a large-scale replication study of TCS techniques in the context of LLM code models. We re-examine established TCS strategies originally proposed for DNNs and complement them with statistical sampling strategies that have not previously been evaluated for TCS. We assess their effectiveness on three code-related classification tasks: clone detection, vulnerability detection, and technical debt prediction. 
\textcolor{black}{The study spans 17 task-specific fine-tuned model instances, 7 predictive features, and 13 selection strategies, including 12 feature-aware strategies and simple random sampling (SRS) as a feature-agnostic baseline. We evaluate performance along two dimensions: operational accuracy estimation and early failure discovery.}

\textcolor{black}{The results indicate that only a subset of findings reported for vision-based DNNs generalize when TCS is applied to LLMs for code. In particular, uncertainty-based features are effective for early failure discovery, while representation-based features are more robust for accuracy estimation.} At the same time, performance varies substantially across tasks and models, indicating that the effectiveness of TCS techniques is context-dependent. Overall, this study provides empirical evidence on the replicability of TCS techniques beyond vision-based deep learning and offers insights into their use for the operational evaluation of LLMs for code.
\end{abstract}



\maketitle

\begin{center}

\textit{Accepted at the 35th ACM SIGSOFT International Symposium on Software Testing and Analysis (ISSTA 2026).}

\end{center}
\section{Introduction}

Test case selection (TCS) is a well-established research area that addresses the challenge of evaluating systems under limited testing budgets~\cite{hu2024test}. Rather than exhaustively inspecting all available test cases, TCS techniques prioritize which cases to examine to maximize testing effectiveness. In traditional software testing, TCS approaches have been developed based on criteria such as coverage~\cite{yoo2012regression}, historical faults~\cite{mondal2015exploring}, and diversity~\cite{noor2015similarity}. More recently, TCS has been introduced in the context of deep neural networks (DNNs)~\cite{hu2024test}, with techniques such as uncertainty-based prioritization~\cite{asgari2025adaptive,guerriero2024deepsample}, surprise adequacy~\cite{kim2019guiding}, and adaptive sampling~\cite{gao2022adaptive}.

These DNN-focused TCS techniques are typically formulated in model-agnostic terms and are discussed as generally applicable to deep learning models. However, empirical evaluations have predominantly focused on vision-based classification tasks. A recent survey by Hu et al.~\cite{hu2024test} explicitly observes that existing software engineering studies on DNN testing do not consider datasets or models specifically designed for software engineering tasks or artifacts and instead rely exclusively on datasets and models originally provided by the machine learning community. As a result, despite promising reported results, it remains unclear to what extent conclusions drawn from prior TCS studies generalize to models for code.

Large language models (LLMs) for code fall within the class of deep neural networks targeted by prior DNN testing research. At a high level, both vision models and code models operate under a supervised learning paradigm and are evaluated on classification or prediction tasks. However, the input modality and task characteristics are quite different. Software artifacts, including source code, are structured inputs: they follow specific linguistic patterns~\cite{capobianco2013improving}, adhere to formal grammars~\cite{sun2024ai}, and exhibit rich semantics~\cite{allamanis2018survey}. Moreover, the cost and expertise required to validate individual test cases can vary widely across inputs. These characteristics make the evaluation of code models particularly challenging.

LLMs for code are increasingly adopted for a wide range of software engineering tasks, including vulnerability detection~\cite{zhou2024large, chen2023diversevul}, clone detection~\cite{khajezade2024investigating, sonnekalb2022generalizability}, and technical debt analysis~\cite{astekin2025detecting}. As these models move from research prototypes to practical tools, understanding their reliability in realistic operational settings has become an important concern~\cite{asgarimetamorphic}. Yet evaluating LLM-based code models at scale remains difficult. Inference is expensive due to token-based computation costs~\cite{guo2024stop}, and assessing correctness often requires expert judgment, as automated oracles are rarely available~\cite{barr2014oracle}. For example, when LLMs are applied to unseen code, validating their outputs may require a domain expert, such as a security analyst in vulnerability detection tasks.
\textcolor{black}{However, when full benchmark evaluation is infeasible, evaluation is limited to a subset of test cases, and simple subsampling provides limited guidance on which inputs should be inspected under constrained validation budgets. Each test case consists of a code artifact (e.g., a code snippet or documentation) paired with a reference label provided by a domain expert~\cite{svajlenko2021bigclonebench}, and a failure occurs when the model's output does not match this label. The choice of which code artifacts to inspect directly affects how many model failures can be identified within a given evaluation budget. As a result, evaluating LLMs for code under realistic constraints can be viewed as a TCS problem.}

In this paper, we present a systematic replication study that investigates the generalizability of existing TCS approaches for DNNs in the context of code-related LLM applications. Rather than proposing new TCS techniques, our goal is to assess whether established findings—derived largely from vision-based evaluations—replicate when applied to expert-driven evaluation of code models.

We study test case selection as a combination of predictive features used to characterize test cases and selection strategies that prioritize them under a limited inspection budget. Specifically, we evaluate seven predictive features derived from model outputs and internal representations, and apply thirteen selection strategies in total. These include seven strategies previously studied in the context of traditional software systems or image-based DNNs, as well as six strategies from the statistical sampling literature that, to our knowledge, have not previously been evaluated for TCS. This design allows us to systematically assess the robustness and generalizability of existing TCS techniques when applied to LLMs for code.

\textcolor{black}{
We conduct a large-scale evaluation across three classification-based code-related tasks: vulnerability detection~\cite{zhou2024large}, clone detection~\cite{khajezade2024investigating}, and technical debt analysis~\cite{astekin2025detecting}. This focus is consistent with common software-engineering uses of code models and LLMs for classification and prediction tasks~\cite{hou2023large,zhang2023survey}. Our study covers 17 task-specific fine-tuned model instances derived from 12 open-source base models, including CodeLLaMA and CodeT5+, under multiple selection budgets. We consider two complementary objectives: (1) operational accuracy estimation under limited evaluation budgets, and (2) early failure discovery, measuring how effectively failures are identified as inspection progresses.
We focus on classification tasks to preserve methodological comparability with prior DNN TCS studies~\cite{hu2024test,guerriero2024deepsample,asgari2025adaptive}, which evaluated these techniques on vision-based classification problems. This controlled scope allows us to examine whether prior TCS findings transfer to LLM-based code classification tasks and which findings do or do not carry over. Extending these techniques to generative code tasks is non-trivial because outputs are sequences rather than single labels, and existing confidence, uncertainty, and representation-based features do not directly map to token-level or sequence-level generation outputs. Since code generation is a major LLM4Code setting~\cite{chen2023large,zheng2023surveycode,fan2023large}, adapting TCS to generative outputs remains future work.
}

Our results indicate that only a subset of findings reported for vision-based DNNs replicate when test case selection is applied to LLMs for code. Across tasks and models, the effectiveness of TCS techniques varies substantially, suggesting that their performance is highly context-dependent. In particular, feature-driven uncertainty measures consistently outperform entropy-based baselines in terms of early failure discovery, while exhibiting more stable behavior across evaluation budgets. At the same time, statistical sampling–based strategies improve the robustness of performance estimation and do not systematically trade off accuracy estimation against failure discovery.

Overall, these results show that while several core insights from prior DNN TCS studies generalize to code-related LLM tasks, others do not. Rather than indicating that existing techniques are ineffective, our findings highlight the importance of understanding which classes of TCS strategies replicate under different task and model characteristics, and how they can be applied to support the operational evaluation of LLMs for code.
The contributions of this paper are as follows:
\begin{itemize}[topsep=2pt, itemsep=1pt]
    \item \textcolor{black}{We present a systematic replication study of test case selection (TCS) techniques for deep neural networks,  evaluating their applicability to LLM-based code classification tasks.}
    \item We empirically assess which findings from prior DNN testing studies generalize to code-related tasks, and which are sensitive to task, model, or evaluation objective.
    \item \textcolor{black}{We provide a large-scale experimental benchmark across multiple code-related tasks, models, predictive features, and selection strategies, offering practical guidance for operational evaluation of LLMs on code classification tasks under limited testing budgets.}
\end{itemize}

\section{Background and Related Work}
\label{sec:background}

Test case selection (TCS) is a central problem in the broader area of test
optimization for deep neural networks, whose goal is to reduce testing and
labeling costs while preserving the effectiveness of model evaluation.
According to the recent survey by Hu et al.~\cite{hu2025assessing}, test
optimization techniques for deep models are commonly organized around four
testing goals: fault detection, performance estimation, model selection,
and sampling-based model retraining. Among these, fault detection and performance estimation are the two objectives
most directly related to operational evaluation under limited selection budgets~\cite{hu2025assessing},
and they form the focus of this study.

\textcolor{black}{The TCS techniques replicated in this study operate in a supervised classification 
setting and rely on auxiliary variables (or features) derived from model outputs 
or internal representations. }
Common examples include uncertainty-based measures such as
confidence, entropy, margin, and DeepGini~\cite{feng2020deepgini}, as well as
representation-based distances such as surprise adequacy
(DSA/LSA/MDSA)~\cite{kim2019guiding}. Given a fixed
budget, TCS strategies either prioritize inputs that are likely to be
misclassified, supporting early failure discovery, or select representative
subsets that allow accurate estimation of model performance on the full test
set (or test suite) using statistical estimators~\cite{lohr2021sampling}. These two
objectives are complementary but often conflicting, and prior work consistently
reports that different strategies perform best depending on which objective is
emphasized.

Empirical evaluations of TCS techniques have so far been dominated by
vision-based classification tasks, typically using convolutional DNNs and image datasets. Large-scale studies such as DeepSample
systematically compare classical sampling designs and adaptive strategies,
showing that estimation-oriented methods (e.g., stratified and
unequal-probability sampling) tend to perform well for accuracy estimation,
while uncertainty-driven and adaptive strategies are more effective for early
failure discovery~\cite{guerriero2024deepsample}. Similar conclusions are
reported in other image-based studies on test prioritization and adaptive test
selection~\cite{gao2022adaptive,ma2021test,sun2023robust}. However, as also
highlighted in prior surveys~\cite{hu2025assessing}, these conclusions are drawn exclusively from non-software engineering artifacts, leaving open the
question of whether they generalize to code-related tasks.

\textcolor{black}{
Code-related tasks differ from image-based settings because they involve structured textual artifacts with syntactic and semantic constraints, and correctness assessment often requires domain expertise. These differences may affect how reliably predictive features identify failure-prone inputs and whether selection strategies remain effective for accuracy estimation and failure discovery. Therefore, it is not sufficient to assume that rankings observed on image-based DNNs directly transfer to LLM-based code classification tasks.
}

More recently, a small number of studies have explored test selection for LLMs in natural language processing tasks, such as sentiment
analysis~\cite{asgari2025adaptive}. These works confirm that TCS effectiveness
remains objective-dependent, but they typically evaluate a single model and a
single task, and do not consider software engineering tasks or
heterogeneous model families tailored for code.

This study is a replicability assessment of prior empirical results on TCS for
deep models. As an independent research team, we replicate existing evaluations
under a comparable protocol, but in a different application context and with a
broader experimental setup, to assess what generalizes to code-related tasks. The next section describes our replication methodology.

\section{Replication Methodology}
\label{sec:methodology}



This section describes the experimental protocol of our replication study.
Since our primary objective is to assess the generalizability of prior empirical
findings on test case selection, we structure the study around research
questions that mirror the core evaluation objectives used in earlier work~\cite{hu2024test}.
Accordingly, our investigation is guided by the following research questions:

\begin{itemize}
    \item \textbf{RQ1:} \textit{How effective are TCS techniques at estimating the operational accuracy of LLMs for code under limited testing budgets?}  

    \item \textbf{RQ2:} \textit{How effective are TCS techniques at prioritizing failing test cases when applied to LLMs for code under limited testing budgets?}  

\end{itemize}

\subsection{Replication Scope and Design Choices}
\label{sec:scope}

To support a controlled replication, we clarify which aspects
of prior test case selection studies are preserved in our experimental design
and which aspects are intentionally changed.

\textbf{Replicated elements.}
We preserve the core methodological assumptions and evaluation protocol adopted
in prior work on TCS for deep models~\cite{hu2024test}.
In particular, (i) test case selection is performed under a fixed inspection
budget, (ii) selection strategies rely exclusively on auxiliary risk variables (features) derived from model outputs or internal representations, (iii) failures are
defined as mis-predictions with respect to ground-truth labels, and (iv)
effectiveness is evaluated along two complementary dimensions: operational
performance estimation and early failure discovery.
Importantly, we also preserve the high-level task formulation adopted in prior
studies: all considered problems are formulated as supervised classification tasks.
While prior work focuses on image-based classification (including binary and
multi-label object recognition), our study considers binary (e.g., vulnerability detection) and multi-label
classification tasks (e.g., technical debt prediction).

We deliberately exclude generative tasks (e.g., code generation) from this study. Generative settings fundamentally alter both the definition of failures and the nature of predictive features, requiring different evaluation metrics that are not directly comparable to those used in classification.
Restricting our scope to classification tasks allows us to preserve methodological comparability with prior TCS studies and to isolate the effect of the application domain and model family. Extending test case selection to generative LLM tasks remains an important direction for future work.

\textbf{Difference from prior studies.}
While preserving the above elements, our study departs from prior work along
several controlled dimensions.
First, we change the \emph{input domain} from images to source code, introducing
textual inputs with different semantic properties~\cite{allamanis2018survey}.
Second, we focus on \emph{transformer-based deep models}, rather than
convolutional deep models used in vision-based
benchmarks.
Third, we evaluate test case selection on \emph{software engineering tasks} with
distinct failure semantics, including binary and multi-label classification, and
tasks whose errors are directly associated with software quality, security, and
maintainability.
Finally, we consider a heterogeneous set of models that vary substantially in
size, architecture, and pre-training objectives, reflecting the diversity of
LLMs used in software engineering tasks~\cite{zhang2023survey}.

\subsection{Tasks, Datasets, and Models}
\label{sec:tasks_models}

To assess the generality of prior TCS techniques, we evaluate them on three widely studied software engineering tasks: clone detection~\cite{svajlenko2021bigclonebench}, vulnerability detection~\cite{zhou2019devign}, and technical debt prediction~\cite{tesoro-code-dataset}. All tasks are formulated as supervised classification problems and are commonly used benchmarks for evaluating LLMs for code~\cite{koohestani2025benchmarking, ni2026learning, le2025impacts}.

\subsubsection{Downstream Tasks}
\label{sec:tasks}
\noindent \textit{Clone detection} aims to determine whether two code fragments implement
the same functionality. We use the BigCloneBench dataset~\cite{svajlenko2021bigclonebench},
a large-scale benchmark comprising over 25{,}000 open-source Java projects from
\texttt{SourceForge} and \texttt{Google Code}. BigCloneBench is a standard dataset for evaluating clone detection models and has been widely used in prior work on neural code understanding~\cite{koohestani2025benchmarking}. The task is formulated as binary classification, where each instance consists of a pair of code snippets and the model predicts whether they are semantic clones.

\textit{Vulnerability detection} focuses on identifying security vulnerabilities
in given code snippets. We use the Devign dataset~\cite{zhou2019devign}, which contains C code
snippets mined from the \texttt{FFmpeg} and \texttt{Qemu} projects and labeled
according to the presence of known vulnerabilities. Devign is a widely adopted
benchmark for learning-based vulnerability detection~\cite{ni2026learning}. This
task is also formulated as binary classification, where each input program is
labeled as vulnerable or non-vulnerable.

\textit{Technical debt prediction} aims to identify self-admitted maintainability
issues in source code. We use the Tesoro dataset~\cite{tesoro-code-dataset}, which
contains code snippets from 974 real-world Java projects annotated with multiple
categories of technical debt, such as code smells and design issues. Unlike the
other tasks, technical debt prediction is formulated as a \emph{multi-label}
classification problem, as a single snippet may exhibit multiple types of debt.
Tesoro has been used in prior work on technical debt analysis and
is particularly challenging due to label heterogeneity and class imbalance~\cite{le2025impacts}.

\begin{table}[t]
\centering
\caption{Models evaluated across three software analytics tasks.
For each model we report the dataset, detection type, number of records, and full-population accuracy.}
\label{tab:model_summary}
\scriptsize
\setlength{\tabcolsep}{4pt}
\renewcommand{\arraystretch}{0.95}

\begin{tabular}{l l l c r r}
\toprule
\textbf{Task} & \textbf{Model} & \textbf{Dataset} & \textbf{Task Type} &
\textbf{\#Records} & \textbf{Reported Accuracy} \\
\midrule
\multirow{3}{*}{Clone Detection} & 
CodeBERT~\cite{feng2020codebert}        & \multirow{3}{*}{BigCloneBench~\cite{svajlenko2021bigclonebench}} & Binary & 415{,}416 & 0.7545 \\
& ModernBERT~\cite{warner2025smarter}       &  & Binary & 415{,}416 & \textbf{0.9900} \\
& UniXCoder~\cite{guo2022unixcoder}       &  & Binary & 415{,}416 & 0.6565 \\
\midrule
\multirow{5}{*}{Vulnerability Detection} & CodeBERT~\cite{feng2020codebert}        & \multirow{5}{*}{Devign~\cite{zhou2019devign}} & Binary & 2{,}732 & 0.6409 \\
& GraphCodeBERT~\cite{guo2020graphcodebert}   &  & Binary & 2{,}732 & 0.6428 \\
& ModernBERT~\cite{warner2025smarter}      &  & Binary & 2{,}732 & 0.4597 \\
& StarEncoder~\cite{li2023starcoder}     &  & Binary & 2{,}732 & \textbf{0.6870} \\
& UniXCoder~\cite{guo2022unixcoder}       &  & Binary & 2{,}732 & 0.6545 \\
\midrule
\multirow{9}{*}{Technical Debt}  
& CodeLLaMA--7B~\cite{roziere2023code}   & \multirow{9}{*}{Tesoro~\cite{tesoro-code-dataset}} & Binary      & 1{,}255 & 0.4375 \\
& CodeT5+~\cite{wang2023codet5p}         &  & Multi-label & 1{,}255 & 0.4821 \\
& GraphCodeBERT~\cite{guo2020graphcodebert}   &  & Multi-label & 1{,}255 & 0.7004 \\
& Magicoder--6.7B~\cite{wei2023magicoder}  &  & Binary      & 1{,}255 & \textbf{0.7171} \\
& Phi--2~\cite{javaheripi2023phi}          &  & Multi-label & 1{,}255 & 0.5729 \\
& PLBART~\cite{ahmad2021unified}          &  & Multi-label & 1{,}255 & 0.6805 \\
& StarCoder2~\cite{lozhkov2024starcoder}      &  & Multi-label & 1{,}255 & 0.4191 \\
& TinyLLaMA~\cite{zhang2024tinyllama}       &  & Multi-label & 1{,}255 & 0.6733 \\
& UniXCoder~\cite{guo2022unixcoder}       &  & Multi-label & 1{,}255 & 0.7482 \\
\bottomrule
\end{tabular}
\end{table}

\subsubsection{Large Language Models}
We evaluate a diverse set of transformer-based LLMs that have
been fine-tuned for the three downstream tasks considered in this
study. The selected models span different architectural families, parameter
scales, and pre-training objectives, enabling an assessment of TCS techniques
across heterogeneous code-oriented LLMs.
From an architectural perspective, the model set includes
\emph{encoder-only}, \emph{encoder--decoder}, and \emph{decoder-only}
transformers. 
Encoder-only models (e.g., \texttt{CodeBERT}~\cite{feng2020codebert}, \texttt{GraphCodeBERT}~\cite{guo2020graphcodebert},
\texttt{UniXCoder}~\cite{guo2022unixcoder}, \texttt{ModernBERT}~\cite{warner2025smarter}) are BERT-style architectures primarily
designed for representation learning and discriminative tasks.
Encoder--decoder models (e.g., \texttt{CodeT5+}~\cite{wang2023codet5p}, \texttt{PLBART}~\cite{ahmad2021unified}) follow a
sequence-to-sequence design originally developed for generation but commonly
used for classification via task-specific heads.
Decoder-only models (e.g., \texttt{CodeLLaMA--7B}~\cite{roziere2023code}, \texttt{Magicoder--6.7B}~\cite{wei2023magicoder},
\texttt{StarCoder2}~\cite{lozhkov2024starcoder}, \texttt{Phi--2}~\cite{javaheripi2023phi}, \texttt{TinyLLaMA}~\cite{zhang2024tinyllama}) are autoregressive
transformers adapted to classification through fine-tuning.
The evaluated models vary substantially in size, ranging from hundreds of
millions of parameters (e.g., \texttt{CodeBERT}, \texttt{GraphCodeBERT},
\texttt{UniXCoder}) to multi-billion-parameter models
(e.g., \texttt{CodeLLaMA--7B}, \texttt{Magicoder--6.7B}, \texttt{StarCoder2}).

These models also differ in pre-training objectives.
Encoder-only models are typically trained with masked language modeling over
code and natural language~\cite{feng2020codebert}, encoder--decoder models rely
on denoising or span-corruption objectives~\cite{wang2023codet5p}, and
decoder-only models are pre-trained autoregressively on large-scale code and
text corpora~\cite{roziere2023code}.
Some models further incorporate structural or multimodal signals, such as data
flow information (\texttt{GraphCodeBERT}) or joint code--text objectives
(\texttt{UniXCoder}, \texttt{PLBART}).
Despite these differences, all models return probabilistic outputs suitable for
test case selection, allowing a controlled and comparable replication study.

\subsubsection{Inclusion criteria for the LLMs}
For each task, we include only models for which publicly available checkpoints have been explicitly fine-tuned and evaluated on the corresponding dataset by their original authors or subsequent studies. 
Not all models are used for all tasks; instead, the model–task associations reflect the availability of task-specific fine-tuned checkpoints. Table~\ref{tab:model_summary} summarizes the evaluated models, the datasets on
which they were fine-tuned, the task type, the number of test instances, and the
reported accuracy.

All models are obtained from publicly available checkpoints hosted on \texttt{HuggingFace}\footnote{\url{https://huggingface.co}}. We do not perform additional fine-tuning; instead, we rely on fine-tuned checkpoints for the specific downstream tasks considered that are publicly available. In particular, selected models satisfy the following criteria:
(i) a task-specific fine-tuned checkpoint is available on \texttt{HuggingFace},
(ii) the corresponding studies report performance on a clearly defined test set, and
(iii) a clear separation between training (used for fine-tuning) and test data (used for evaluation) is provided.

\subsection{Test Case Selection Techniques}
\label{sec:tcs}


In this study, we follow the standard decomposition of test case selection into two design components:  (i) the \emph{predictive features} (auxiliary variables) used to characterize test inputs, and  (ii) the \emph{selection strategy} used to prioritize or sample test cases under
a fixed inspection budget.
Given an operational test set, predictive features are computed for each input based on model outputs or internal representations. A selection strategy then uses these features to construct a subset of test cases for inspection.
We adopt this separation to structure our evaluation of test case selection for language models for code. Predictive features are assessed in terms of their relationship to failure occurrence and performance degradation, while selection strategies are evaluated based on their effectiveness and robustness across tasks, models, and testing budgets.

\subsubsection{Predictive Features}
\label{sec:features}

\begin{table}[t]
\centering
\color{black}
\caption{ \textcolor{black}{Predictive features considered in this study. The table summarizes the feature family, intuition, and representative prior sources.}}
\label{tab:feature_summary}
\scriptsize
\setlength{\tabcolsep}{4pt}
\renewcommand{\arraystretch}{1.05}
\begin{tabular}{l l p{7.3cm} l}
\toprule
\textbf{Feature} & \textbf{Family} & \textbf{Main intuition} & \textbf{Representative sources} \\
\midrule
Confidence & Uncertainty-based & Lower confidence indicates higher risk of misprediction. & \cite{asgari2025adaptive,guerriero2024deepsample, hu2024test, wang2016cost} \\
Entropy & Uncertainty-based & Higher predictive entropy indicates greater output ambiguity. & \cite{asgari2025adaptive,hu2024test, wang2016cost} \\
Margin & Uncertainty-based & Smaller gap between top predictions indicates fragile decisions. & \cite{jiang2018predicting, hu2024test, weinstein2020margin, ducoffe2018adversarial, wang2016cost} \\
DeepGini & Uncertainty-based & Flatter predictive distributions indicate higher uncertainty. & \cite{feng2020deepgini,hu2024test, kim2019guiding} \\
LSA & Surprise adequacy & Atypicality relative to the predicted class in representation space. & \cite{kim2019guiding,guerriero2024deepsample, hu2024test, abbasishahkoo2025metasel, aghababaeyan2024deepgd} \\
DSA & Surprise adequacy & Relative input proximity to other-class representations. & \cite{kim2023evaluating,guerriero2024deepsample, hu2024test, kim2019guiding, abbasishahkoo2025metasel, aghababaeyan2024deepgd} \\
MDSA & Surprise adequacy &  Mahalanobis surprise relative to class-conditional representations. & \cite{kim2023evaluating, hu2024test, kim2019guiding, abbasishahkoo2025metasel} \\
\bottomrule
\end{tabular}
\end{table}

\textcolor{black}{
The scope of this study is limited to predictive features that have been used in prior DNN test-selection work and can be extracted consistently across all evaluated models and tasks. Our feature set includes the features used in the DNN-TCS studies of Guerriero et al.~\cite{guerriero2024deepsample} and Asgari et al.~\cite{asgari2025adaptive}, and extends them with additional uncertainty-based and surprise-adequacy features discussed in recent surveys of DNN test optimization~\cite{hu2024test}. This yields a fixed set of seven features: Confidence, Entropy, Margin, DeepGini, LSA, DSA, and MDSA (Table \ref{tab:feature_summary}). Together, they cover two commonly used families of test-selection criteria: output-based uncertainty and representation-based surprise.
}
\textcolor{black}{
In this study, predictive features, also referred to as \emph{auxiliary variables}, are numerical quantities derived from model outputs or internal representations. Selection strategies use these features to rank or sample test cases under a fixed budget, with the goal of prioritizing inputs that are more likely to expose model failures. Importantly, predictive features are used only for prioritization: they do not serve as test oracles and do not affect how prediction correctness is determined.
}
\textcolor{black}{
We do not include transformer-specific internal criteria, such as attention-head-based measures, because they require new design choices, including layer/head selection and aggregation, that are not uniformly comparable across encoder-only, decoder-only, and encoder-decoder models. Evaluating such criteria would constitute a separate extension rather than a controlled replication of existing TCS techniques, so we leave them for future work.
}

Let $\mathbf{p}(x) = (p_1(x), \dots, p_K(x))$ denote the model output for input
$x$ (a code snippet in our setting).
For binary and multi-class tasks, $\mathbf{p}(x)$ is a probability distribution
over $K$ classes.
For multi-label tasks, each $p_k(x)\in[0,1]$ represents the predicted probability
that label $k$ applies to $x$.
To compute auxiliary risk variables consistently across task types, we define a
proxy predicted label as $\hat{y}(x) = \arg\max_k p_k(x)$.
This proxy is used solely for risk scoring and test selection and does not
affect how prediction correctness is evaluated.

\textbf{Confidence}.
Prediction confidence is the probability assigned to the predicted class:
$\mathrm{Conf}(x) = p_{\hat{y}(x)}(x)$.
Low confidence indicates uncertainty in the predicted decision and is therefore
commonly treated as a proxy for failure risk in black-box test selection and
operational testing~\cite{asgari2025adaptive,guerriero2024deepsample}.
In our study, confidence is a \emph{safety-like} feature and is inverted when
converted into a risk score (so that larger values consistently indicate higher
risk).

\textbf{Entropy}.
Entropy captures the overall uncertainty of the predictive distribution:
\\ $\mathrm{H}(x) = -\sum_{c=1}^{K} p_c(x)\log p_c(x)$.
Higher entropy indicates a more ambiguous prediction, with probability mass
spread across multiple classes.
Entropy has been widely used as a black-box prioritization criterion in
operational testing and adaptive evaluation settings~\cite{asgari2025adaptive}.
In our setting, larger entropy is treated as higher risk.

\textbf{Margin}.
The margin measures the separation between the two most likely
classes:
$\mathrm{Margin}(x) = p_{(1)}(x) - p_{(2)}(x)$,
where $p_{(1)}(x)$ and $p_{(2)}(x)$ are the largest and second-largest predicted
probabilities \cite{jiang2018predicting, hu2024test}.
Small margins indicate fragile decisions near the model’s decision boundary and
are commonly associated with higher failure likelihood.
As with confidence, margin is safety-like and is inverted to represent risk.

\textbf{DeepGini}.
DeepGini~\cite{feng2020deepgini} measures the dispersion of the predictive
distribution:
$\mathrm{DeepGini}(x) = 1 - \sum_{c=1}^{K} p_c(x)^2$.
Larger values correspond to flatter (less decisive) output distributions and
higher uncertainty.
DeepGini is a standard baseline in black-box DNN test selection and has been
shown to correlate with mispredictions~\cite{feng2020deepgini}.
In our study, it serves as both an auxiliary risk variable and a reference
feature for comparing more structured sampling strategies.

\textbf{Likelihood-based Surprise Adequacy (LSA)}.
LSA quantifies how atypical an input is relative to the training data of its
predicted class using hidden-layer representations~\cite{kim2019guiding,guerriero2024deepsample}.
Let $z(x)\in\mathbb{R}^d$ be an activation vector for input $x$, and let $f_c$
denote a class-conditional density estimator fitted on training activations for
class $c$.
LSA is defined as $\mathrm{LSA}(x) = -\log f_{\hat{y}(x)}(z(x))$.
Large LSA values indicate low likelihood under the predicted class’s activation
distribution, suggesting operational novelty and potentially higher failure
risk.

\textbf{Distance-based Surprise Adequacy (DSA)}.
DSA measures relative distances to training examples in representation space and
prioritizes inputs that are atypical for their predicted class yet close to
other classes~\cite{kim2023evaluating,guerriero2024deepsample}.
It is defined as
$\mathrm{DSA}(x) =
\frac{d_{\mathrm{same}}(x)}{d_{\mathrm{diff}}(x) + \varepsilon}$,
where $d_{\mathrm{same}}(x)$ is the distance from $z(x)$ to the nearest training
example of the predicted class $\hat{y}(x)$, $d_{\mathrm{diff}}(x)$ is the
distance to the nearest training example from a different class, and
$\varepsilon>0$ prevents division by zero.
Larger DSA values indicate higher surprise and are treated as higher risk.

\textbf{Mahalanobis Distance Surprise Adequacy (MDSA)}.
MDSA models class-conditional activation distributions using Gaussian statistics
and scores how far an input deviates from typical activations of its predicted
class~\cite{kim2019guiding, kim2023evaluating}.
For predicted class $\hat{y}(x)$ with mean $\mu_{\hat{y}}$ and covariance
$\Sigma_{\hat{y}}$, MDSA is defined as
$\mathrm{MDSA}(x) =
(z(x)-\mu_{\hat{y}})^\top \Sigma_{\hat{y}}^{-1}(z(x)-\mu_{\hat{y}})$.
Larger values indicate structurally unusual internal representations and are
treated as higher failure risk.


\begin{table}[t]
\caption{\textcolor{black}{TCS strategies and their design characteristics. 
(*) denotes strategies previously evaluated or adopted in DNN-TCS studies; (**) denotes additional sampling-based strategies from the statistical sampling literature. 
\ding{51} indicates that a property (e.g., partitioning) is satisfied, and \ding{55} otherwise.}}

\centering
\setlength{\tabcolsep}{5pt}
\renewcommand{\arraystretch}{1.10}
\resizebox{0.85\columnwidth}{!}{
\scriptsize
\begin{tabular}{lcccccc}
\toprule
\textbf{Technique} &
\textbf{Partitioning} &
\textbf{Unequal Prob.} &
\textbf{W/o Replacement} &
\textbf{RA-Strat.} &
\textbf{Adaptive Alloc. } &
\textbf{Unbiased Est.} \\
\midrule
SRS*     & \ding{55} & \ding{55} & \ding{55} & \ding{55} & \ding{55} & \ding{51} \\
SUPS*    & \ding{55} & \ding{51} & \ding{55} & \ding{55} & \ding{55} & \ding{51} \\
RHC-S*   & \ding{55} & \ding{51} & \ding{51} & \ding{55} & \ding{55} & \ding{51} \\
SSRS*    & \ding{51} & \ding{55} & \ding{51} & \ding{55} & \ding{55} & \ding{51} \\
GBS*     & \ding{51} & \ding{55} & \ding{55} & \ding{55} & \ding{51} & \ding{51} \\
2-UPS*   & \ding{51} & \ding{51} & \ding{51} & \ding{51} & \ding{55} & \ding{51} \\
DeepEST* & \ding{55} & \ding{51} & \ding{51} & \ding{55} & \ding{51} & \ding{51} \\
\midrule
Balanced PPS** & \ding{55} & \ding{51} & \ding{51} & \ding{55} & \ding{55} & \ding{51} \\
PPSsys**       & \ding{55} & \ding{51} & \ding{51} & \ding{55} & \ding{55} & \ding{51} \\
Pivotal**      & \ding{55} & \ding{51} & \ding{51} & \ding{55} & \ding{55} & \ding{51} \\
D$^2$Strat**   & \ding{51} & \ding{55} & \ding{51} & \ding{51} & \ding{55} & \ding{51} \\
KCenterHT**    & \ding{51} & \ding{55} & \ding{51} & \ding{51} & \ding{55} & \ding{51} \\
MCUCB**       & \ding{51} & \ding{51} & \ding{51} & \ding{51} & \ding{51} & \ding{51} \\
\bottomrule
\end{tabular}
}
\label{tab:compared_testing_techniques}
\end{table}

\vspace{-5pt}
\subsubsection{Test Selection Strategies}
\label{sec:strategies}

We evaluate thirteen TCS strategies in total. 
\textcolor{black}{
Our goal is not to cover every possible TCS method, but to evaluate a representative and reproducible set of strategies under a shared protocol. To ensure comparability, we include only strategies that can be applied consistently in the same budgeted operational-testing setting. All evaluated strategies except SRS use the predictive features defined in Section~\ref{sec:features}; SRS serves as a feature-agnostic baseline.
}

\textcolor{black}{
Seven strategies are selected from prior software engineering studies on DNN test case selection that evaluated or discussed multiple methods under comparable settings~\cite{guerriero2024deepsample,asgari2025adaptive,hu2024test}. This provides a direct basis for replication. In particular, our study covers all strategies considered by Guerriero et al.~\cite{guerriero2024deepsample} and Asgari et al.~\cite{asgari2025adaptive}, except Cross-Entropy Sampling (CES). We exclude CES because its original formulation constructs samples by matching the distribution of neuron-level representations over the operational dataset, making it less compatible with the shared predictive-feature-driven setup used in this study. This choice is also consistent with the later LLM-focused study by Asgari et al.~\cite{asgari2025adaptive}, which did not include CES.
}
\textcolor{black}{
The remaining six strategies are selected from the statistical sampling literature~\cite{madow1949theory,deville1998unequal,horvitz1952generalization,hajek1971comment,tille2006sampling,carpentier2012adaptive,lohr2021sampling,arthur2006k,arthur2007proceedings,gonzalez1985clustering}. They cover complementary sampling-design properties, including unequal-probability sampling, stratification, diversity-based partitioning, and adaptive allocation. To our knowledge, these strategies have not been evaluated in TCS for DNNs or LLMs. Including them allows us to compare strategies previously evaluated in DNN-TCS studies with general-purpose sampling designs under the same assumptions and evaluation protocol.
}


Table~\ref{tab:compared_testing_techniques} summarizes all test case selection strategies considered in this study and highlights their key design properties. These include whether a strategy relies on partitioning or stratification, uses unequal-probability sampling, samples without replacement, applies risk-aware stratification, performs adaptive allocation, or supports design-based unbiased estimation.
We distinguish between two families of strategies based on their role in this replication. Strategies previously evaluated or adopted in DNN-TCS studies are methods used in prior DNN testing work. Statistical sampling strategies are selected from the statistical sampling literature and have not previously been applied to TCS for DNNs.
\emph{Statistical sampling strategies} are selected from the statistical sampling literature and have not previously been applied to TCS for DNNs.

All strategies are evaluated under the same budgeted operational testing setting. Given an operational dataset $D$ of size $N$=$|D|$ and a fixed testing budget $n$, each strategy selects a subset of $n$ test inputs and is evaluated with respect to two objectives: 
(i) maximizing failure discovery, measured as the number of failures observed within the budget, and 
(ii) accurately estimating the operational failure rate, and hence accuracy, from the selected subset.

Let $y_i \in \{0,1\}$ denote the failure indicator of test input $i$, where $y_i$=$1$ if the model fails on input $i$ and $y_i$=$0$ otherwise. We define the true operational failure rate as
$
\theta = \frac{1}{N}\sum_{i=1}^{N} y_i.
$
All strategies rely on a predictive feature $\chi_i$ (see Section~\ref{sec:features}), derived from model outputs, which is used to guide selection but is never queried as an oracle.

\textbf{Simple Random Sampling (SRS)}.
SRS selects $n$ test cases uniformly at random from $D$ without using any feature $\chi_i$~\cite{pietrantuono2016adaptive,li2019boosting}. Each test case has the same probability of being selected. SRS serves as a baseline and yields an unbiased estimator of the operational failure rate.

\textbf{Simple Unequal Probability Sampling (SUPS)}.
SUPS uses a feature score $\chi_i \geq 0$, derived from model outputs, to perform probability-proportional-to-size sampling with replacement~\cite{lohr2021sampling}. At each draw, test input $i$ is selected with probability
$\pi_i = \chi_i / \sum_{j=1}^{N} \chi_j.$
Test cases with larger feature values are therefore more likely to be selected. To estimate the operational failure rate under biased sampling, SUPS applies the Hansen--Hurwitz estimator~\cite{hansen1943theory}.

\textbf{RHC-Sampling (RHC-S)}.
RHC-Sampling uses a feature $\chi_i \geq 0$ to perform unequal-probability sampling without replacement, prioritizing test cases with larger auxiliary values~\cite{lohr2021sampling}. Selection probabilities are derived from $\chi_i$, but once a test case is selected it is not eligible for subsequent draws. Compared to sampling with replacement, this strategy avoids repeated selection of the same inputs while still biasing the sample toward higher-risk test cases. Estimation of the operational failure rate accounts for unequal inclusion probabilities using a design-based estimator.

\textbf{DeepEST}.
DeepEST is an adaptive test case selection strategy originally proposed for operational testing of deep neural networks~\cite{guerriero2021operation}. The strategy incrementally selects test cases based on an auxiliary variable $\chi$, while balancing exploitation and exploration. At each step, unselected test cases are sampled with probabilities that depend on their similarity to previously selected inputs, as measured by weights derived from $\chi$, and a uniform exploration component. Because selection probabilities change over time, estimation of the operational failure rate requires correction for adaptive inclusion probabilities using a design-based estimator~\cite{horvitz1952generalization}.

\textbf{Stratified Simple Random Sampling (SSRS)}.
SSRS partitions the operational dataset $D$ into $P$ disjoint strata based on the auxiliary variable $\chi$ and applies simple random sampling within each stratum. The number of samples allocated to each stratum is determined using Neyman allocation~\cite{lohr2021sampling}, with the total number of selected test cases fixed to $n$. Let $N_p$ denote the size of stratum $p$ and $n_p$ the number of samples drawn from that stratum. The operational failure rate is estimated using the standard stratified estimator,
$
\hat{\theta} = \sum_{p=1}^{P} \frac{N_p}{N} \cdot \frac{s_p}{n_p},
$,  where $s_p$ denotes the number of observed failures among the sampled test cases in stratum $p$.

\textbf{Gradient-Based Sampling (GBS)}.
GBS is an adaptive stratified sampling strategy that dynamically allocates samples across strata to reduce the variance of the failure-rate estimator~\cite{lohr2021sampling}. At each step, the strategy selects the stratum from which to draw the next test case based on the current stratum sizes and empirical variance estimates. Specifically, let $n_p(t-1)$ denote the number of samples already drawn from stratum $p$ before step $t$, and let $\widehat{V}_p(t-1)$ be the estimated variance of the failure indicator within that stratum. GBS selects the next stratum as:
$
p_t=\arg\max_{p\in\{1,\dots,P\}}
\left(\frac{N_p}{N}\right)^2
\frac{\widehat{V}_p(t-1)}{(n_p(t-1)+1)^2}
$, 
and then draws one additional test case uniformly at random from the selected stratum.

\textbf{Two-Stage Unequal Probability Sampling (2-UPS)}.
2-UPS performs test selection in two stages~\cite{guerriero2024deepsample}.
First, a stratum is selected with probability proportional to the aggregate risk of its elements.
Second, a test input is sampled uniformly from the chosen stratum. This design biases selection toward high-risk regions while maintaining simplicity.
An unbiased estimate of the operational failure rate is obtained using a Hansen--Hurwitz estimator.

\textbf{PPS-Systematic (PPSsys)}.
PPSsys is a fixed-size unequal-probability sampling design without replacement, originally proposed in the survey sampling literature~\cite{madow1949theory,deville1998unequal}.
Each test input $i$ is assigned a non-negative risk score $r_i$, derived from the predictive feature $\chi_i$ and normalized such that $\sum_{i=1}^{N} r_i = 1$. First-order inclusion probabilities are then defined as $\pi_i = \min(1, n \cdot r_i)$, ensuring that the expected sample size is $n$.
A sample $S$ of exactly $n$ test inputs is drawn using systematic PPS without replacement.
The operational failure rate is estimated using the Horvitz--Thompson estimator
\cite{horvitz1952generalization}:
$
\hat{\theta}=\frac{1}{N}\sum_{i\in S}\frac{y_i}{\pi_i}
$.
PPSsys has not previously been applied to TCS for deep models. We include it because it provides a principled baseline for risk-aware, fixed-size sampling with unbiased estimation under a finite testing budget~\cite{madow1949theory,deville1998unequal}.

\textbf{Balanced PPS-Systematic (Balanced)}.
Balanced uses the same fixed-size PPS systematic sampling design and first-order
inclusion probabilities $\{\pi_i\}$ as PPSsys, but differs in how the operational
failure rate is estimated.
Instead of the Horvitz--Thompson estimator, Balanced applies the Hájek
(self-normalized) estimator~\cite{hajek1971comment}:
$
\hat{\theta} =
\frac{\sum_{i\in S} y_i / \pi_i}{\sum_{i\in S} 1 / \pi_i}
$.
The Hájek estimator trades a small amount of bias for substantially reduced variance and improved numerical stability when inclusion probabilities are
highly unequal.
Balanced has not previously been evaluated in the context of TCS for deep models. We include it as a variance-stabilized counterpart to PPSsys, allowing us to assess whether estimator choice affects the robustness of risk-aware sampling.

\textbf{Pivotal Sampling (Pivotal)}.
Pivotal sampling draws an exact fixed-size unequal-probability sample without
replacement using the Deville--Tillé pivotal algorithm~\cite{deville1998unequalSurvey,tille2006sampling}.
Given first-order inclusion probabilities $\pi_i \propto r_i$ derived from the
feature-based risk scores and satisfying $\sum_i \pi_i$=$n$, the algorithm
iteratively updates pairs of units, driving one toward inclusion and the other
toward exclusion while preserving the fixed sample size constraint.
We estimate the operational failure rate using the Hájek (self-normalized)
estimator~\cite{hajek1971comment}, which improves numerical stability under
highly unequal inclusion probabilities.
While pivotal sampling is well established in survey statistics, it has not
previously been applied to test case selection for deep neural networks.
We include it as a principled fixed-size alternative to PPS systematic sampling,
allowing us to assess whether enforcing exact inclusion probabilities affects
failure discovery and accuracy estimation in LLM-based testing.

\textbf{Monte Carlo Upper Confidence Bound (MCUCB)}.
MCUCB is an adaptive stratified sampling strategy originally proposed for Monte Carlo integration \cite{carpentier2012adaptive}. It maintains a fixed set of strata and allocates samples adaptively based on upper confidence bounds computed from the observed failure rates within each stratum. Strata with higher estimated failure rates and higher uncertainty receive more samples, balancing exploration and exploitation. After the budget is exhausted, the operational failure rate is estimated using the standard stratified estimator \cite{lohr2021sampling,tille2006sampling}.
%
While MCUCB has been studied in the context of adaptive Monte Carlo estimation, it has not previously been applied to TCS for deep models. We include MCUCB to assess whether bandit-style adaptive allocation improves early failure discovery and accuracy estimation when strata are defined using feature-based risk signals for LLMs on code-related tasks.

\textbf{D$^2$-Stratified Sampling (D$^2$Strat)}.
D$^2$Strat is a risk-aware stratified sampling strategy that constructs strata
by applying the $k$-means++ (D$^2$) seeding heuristic~\cite{arthur2006k,arthur2007proceedings}
to a one-dimensional auxiliary risk score derived from predictive features.
The seeding step is used solely to obtain a diverse and well-spread partition of
the risk space, without performing iterative clustering.
After selecting $H$ centers, each test input is assigned to its nearest center,
forming $H$ disjoint strata. Sampling proceeds via simple random sampling within
each stratum, and estimation uses the standard stratified estimator~\cite{lohr2021sampling}.
D$^2$Strat has not previously been studied for test case selection in deep models.
We include it to assess whether diversity-driven stratification of risk scores
improves failure discovery and accuracy estimation under limited budgets.

\textbf{$K$-Center Stratified Sampling (KCenterHT)}.
KCenterHT is a risk-aware stratified sampling strategy that constructs strata
using the farthest-first $k$-center heuristic~\cite{gonzalez1985clustering}
applied to a one-dimensional auxiliary risk score derived from predictive
features. Unlike probabilistic seeding, the $k$-center heuristic deterministically
selects centers to maximize coverage of the risk space, ensuring that both
extreme and intermediate risk regions are represented.
After selecting $H$ centers, test inputs are assigned to their nearest center,
forming $H$ disjoint strata. Sampling is performed via simple random sampling
within each stratum, and estimation uses the standard stratified estimator~\cite{lohr2021sampling}.
KCenterHT has not previously been studied for test case selection in deep models.
We include it to contrast deterministic, coverage-oriented stratification with
probabilistic approaches (e.g., D$^2$Strat), and to assess whether explicit
risk-space coverage improves failure discovery and estimation stability under
budgeted testing.

\subsection{Evaluation Metrics}
\label{sec:evaluation_methodology}

This section describes how test case selection techniques are applied to the models and tasks introduced in Section~\ref{sec:tasks_models}, and how their effectiveness is evaluated with respect to the research questions defined in Section~\ref{sec:methodology}.
%
For each model-task pair, we treat the corresponding test split as an \emph{operational dataset} $D$ of size $N$. Given a predictive feature $\chi_i$ computed for each test input $i \in D$, a test selection strategy selects a subset of $n$ inputs under a fixed budget. Ground-truth labels are used \emph{only} for evaluation and never for guiding
selection. 

\textit{Experimental setup}.
All experiments are conducted over $R=30$ independent runs. Across all tasks, we evaluate $17$  task-specific fine-tuned model instances using $12$ feature-aware TCS strategies and $7$ predictive features.
In addition, Simple Random Sampling (SRS) is included as a non-risk-aware baseline that does not rely on predictive features.
For \textbf{RQ1} (accuracy estimation) and \textbf{RQ2} (early failure discovery),
we report results for a fixed test selection budget of $n=200$, following prior
work on operational testing of deep models, where budgets in this range are
commonly used to balance labeling cost and statistical reliability
\cite{guerriero2021operation,guerriero2024deepsample,asgari2025adaptive}.
\textcolor{black}{We use fixed absolute budgets rather than proportional budgets because they reflect a practitioner-facing scenario in which the limiting resource is expert inspection effort rather than the size of the available operational dataset.}
We additionally evaluate all techniques under multiple inspection budgets
$n \in \{50,100,200,400,800\}$.
Due to page limits, detailed results are reported only for $n=200$; results for
other budgets exhibit consistent qualitative patterns, which are summarized in
Section~\ref{sec:discussion}. The complete results are available in the
replication package.
In total, we performed
$
    17 \textrm{ (models)} \times 30 \textrm{ (repetitions)} \times 
    \left[
    12 \textrm{ (strategies)} \times 7 \textrm{ (features)} + 1 \textrm{ (SRS)}
    \right]
    = 43{,}350
$
experiments per testing budget.
%

\textit{Accuracy estimation (RQ1)}.
For each model--task combination, we compute the \emph{true operational accuracy}
by evaluating the model on the full operational test set.
This accuracy serves as ground truth and is never used to guide test selection.
Given a test selection strategy and a predictive feature, a subset of $n$ test
inputs is selected from the same test set and used to compute an accuracy
estimate.
This procedure is repeated for $R$ independent runs with different sampled
subsets.
Estimation quality is measured using the root mean squared error (RMSE) between
the sampled accuracy estimates and the true operational accuracy~\cite{hu2024test}.
Lower RMSE indicates more accurate estimation under the given testing budget.

\textit{Early discovery of failing test cases (RQ2)}. 
To evaluate failure discovery, we analyze how effectively a test selection strategy concentrates the testing budget on inputs that the model misclassifies. For a given model, predictive feature, and TCS strategy, the strategy selects $n$ test inputs from the operational test set. The model is executed on each selected input, and a \emph{failure} is recorded whenever the predicted label differs from the ground-truth label. This definition of failure is consistent with prior work on DNN testing, where failures correspond to mispredictions with respect to known labels.
Failure discovery performance is measured as the number of failing test cases observed within the selected budget. Higher values indicate that the strategy is more effective at prioritizing inputs on which the model behaves incorrectly, supporting early debugging, model analysis, and risk assessment under limited testing resources~\cite{hu2024test}.


\textit{Statistical analysis}.
For each combination of model, task, predictive feature, test selection strategy,
and inspection budget $n \in \{50,100,200,400,800\}$, results are aggregated over
$R=30$ independent runs using the median, which is robust to outliers introduced
by stochastic sampling.
For statistical comparison, we aggregate results across all evaluated budgets,
tasks, and models to identify patterns across different testing budgets.
To compare multiple test selection strategies across these settings, we
employ the Friedman test with significance level $\alpha=0.05$~\cite{Garcia:2009}.
The Friedman test is a non-parametric significance test designed to compare multiple
treatments across multiple problem instances using ranking; in
our case, strategies correspond to treatments, while model--task--budget combinations are problem instances.
When the Friedman test indicates statistically significant differences, we conduct post-hoc analysis using the Nemenyi test to identify which strategies
differ significantly from one another~\cite{japkowicz2011evaluating}. Two treatments are significantly different when the difference in their average
ranks exceeds the Nemenyi critical distance.
This combination of Friedman and post-hoc tests is widely used for comparing
randomized algorithms across multiple problems and has been adopted in
search-based software engineering~\cite{devroey2023juge,panichella2021systematic} and benchmarking studies~\cite{japkowicz2011evaluating}.
\vspace{-12pt}

\newcommand{\better}[1]{\textcolor{black}{\textbf{#1}}}
\begin{table}[t]
\centering
\caption{\textcolor{black}{RQ1: Accuracy estimation error (RMSE; median/IQR) for all three tasks at $n=200$. Darker/lighter shading marks the best/second-best strategy for each feature; bold strategy names are new TCS techniques, and bold values outperform the \textit{SRS} baseline.}}

\label{tab:rq1-clone-RMSE}
\scriptsize
\resizebox{.85\textwidth}{!}{
\begin{tabular}{lccccccc}
\bottomrule
\multicolumn{8}{c}{ Code Clone} \\
\toprule
Strategy & Confidence & DeepGini & DSA & LSA & Margin & MDSA & Entropy \\
\toprule
2UPS & 0.077 (0.095) & 0.077 (0.101) & \better{0.020 (0.011)} & 0.028 (0.020) & 0.074 (0.094) & 0.028 (0.023) & 0.084 (0.111) \\
DeepEST & 0.056 (0.039) & 0.056 (0.044) & 0.056 (0.036) & 0.052 (0.032) & 0.049 (0.037) & 0.059 (0.039) & 0.057 (0.045) \\
GBS & \cellcolor{gray!20} \better{0.022 (0.014)} & 0.026 (0.020) & \cellcolor{gray!40} \better{0.019 (0.012)} & 0.026 (0.018) & \better{0.024 (0.016)} & 0.025 (0.017) & 0.030 (0.023) \\
RHCS & 0.091 (0.064) & 0.088 (0.061) & 0.094 (0.075) & 0.085 (0.072) & 0.111 (0.086) & 0.099 (0.062) & 0.113 (0.085) \\
SSRS & 0.047 (0.039) & 0.057 (0.037) & \cellcolor{gray!40} \better{0.019 (0.012)} & \cellcolor{gray!40} \better{0.020 (0.013)} & 0.050 (0.036) & \cellcolor{gray!20} \better{0.023 (0.016)} & 0.047 (0.033) \\
SUPS & 0.026 (0.017) & 0.027 (0.019) & \better{0.022 (0.013)} & 0.025 (0.018) & \cellcolor{gray!40} \better{0.022 (0.015)} & \better{0.024 (0.016)} & \cellcolor{gray!20} 0.027 (0.018) \\
\textbf{Balanced} & 0.068 (0.089) & 0.063 (0.075) & 0.030 (0.022) & 0.034 (0.023) & 0.068 (0.089) & 0.031 (0.024) & 0.053 (0.063) \\
\textbf{D2Strat} & \better{0.024 (0.017)} & \cellcolor{gray!20} \better{0.023 (0.014)} & \cellcolor{gray!20} \better{0.021 (0.011)} & \cellcolor{gray!20}\better{0.022 (0.015)} & \cellcolor{gray!40} \better{0.022 (0.014)} & \cellcolor{gray!40} \better{0.020 (0.011)} & \cellcolor{gray!40} \better{0.022 (0.015)} \\
\textbf{KCenterHT} & \cellcolor{gray!40} \better{0.019 (0.013)} & \cellcolor{gray!40} \better{0.021 (0.013)} & \cellcolor{gray!20} \better{0.021 (0.013)} & \cellcolor{gray!40} \better{0.020 (0.012)} & \cellcolor{gray!40}\better{0.022 (0.015)} & \cellcolor{gray!20} \better{0.023 (0.017)} & \cellcolor{gray!40} \better{0.022 (0.014)} \\
\textbf{MCUCB} & 0.026 (0.016) & 0.027 (0.017) & 0.027 (0.018) & 0.029 (0.020) & 0.026 (0.017) & 0.025 (0.017) & 0.026 (0.016) \\
\textbf{PPSsys} & 0.048 (0.049) & 0.349 (0.568) & \cellcolor{gray!40} \better{0.019 (0.012)} & 0.031 (0.022) & 0.048 (0.050) & 0.035 (0.032) & 0.053 (0.056) \\
\textbf{Pivotal} & 0.137 (0.117) & 0.156 (0.131) & 0.189 (0.163) & 0.117 (0.101) & 0.143 (0.125) & 0.116 (0.100) & 0.132 (0.106) \\
\cline{2-8}
\textit{SRS} & \multicolumn{7}{c}{\textit{0.025 (0.015)}} \\
\bottomrule
\multicolumn{8}{c}{ Technical Debt} \\
\toprule
2UPS & 0.110 (0.068) & 0.105 (0.072) & 0.053 (0.041) & 0.047 (0.022) & 0.081 (0.055) & 0.044 (0.030) & 0.073 (0.045) \\
DeepEST & 0.082 (0.021) & 0.080 (0.015) & 0.083 (0.013) & 0.088 (0.016) & 0.081 (0.016) & 0.082 (0.014) & 0.049 (0.023) \\
GBS & 0.035 (0.005) & \cellcolor{gray!20} \better{0.031 (0.004)} & \better{0.024 (0.011)} & \better{0.028 (0.011)} & 0.035 (0.005) & \better{0.029 (0.008)} & 0.034 (0.006) \\
RHCS & 0.117 (0.036) & 0.121 (0.032) & 0.132 (0.029) & 0.129 (0.027) & 0.105 (0.030) & 0.123 (0.025) & 0.061 (0.040) \\
SSRS & 0.044 (0.014) & 0.044 (0.013) & \cellcolor{gray!20} \better{0.022 (0.008)} & 0.074 (0.101) & 0.044 (0.011) & \cellcolor{gray!20} \better{0.027 (0.005)} & 0.044 (0.017) \\
SUPS & \cellcolor{gray!20} \better{0.032 (0.006)} & 0.035 (0.004) & \better{0.031 (0.007)} & \better{0.030 (0.004)} & \better{0.033 (0.009)} & 0.034 (0.008) & \better{0.031 (0.005)} \\
\textbf{Balanced} & 0.093 (0.047) & 0.076 (0.044) & 0.104 (0.078) & 0.058 (0.030) & 0.087 (0.061) & 0.058 (0.025) & 0.073 (0.035) \\
\textbf{D2Strat} & \cellcolor{gray!20} \better{0.032 (0.005)} & \better{0.032 (0.003)} & \cellcolor{gray!20} \better{0.022 (0.008)} & \cellcolor{gray!40} \better{0.024 (0.005)} & \cellcolor{gray!20} \better{0.031 (0.003)} & \cellcolor{gray!20} \better{0.027 (0.007)} & \cellcolor{gray!40} \better{0.028 (0.005)} \\
\textbf{KCenterHT} & \cellcolor{gray!40} \better{0.028 (0.003)} & \cellcolor{gray!40} \better{0.030 (0.005)} & \cellcolor{gray!40} \better{0.021 (0.008)} & \cellcolor{gray!20} \better{0.025 (0.007)} & \cellcolor{gray!40} \better{0.029 (0.004)} & \cellcolor{gray!40} \better{0.025 (0.004)} & \cellcolor{gray!20} \better{0.029 (0.003)} \\
\textbf{MCUCB} & 0.044 (0.009) & 0.044 (0.010) & 0.040 (0.019) & 0.038 (0.020) & 0.042 (0.011) & 0.038 (0.019) & 0.040 (0.011) \\
\textbf{PPSsys} & 0.094 (0.063) & 0.084 (0.063) & 0.093 (0.093) & 0.056 (0.060) & 0.096 (0.052) & 0.060 (0.060) & 0.061 (0.035) \\
\textbf{Pivotal} & 0.222 (0.119) & 0.224 (0.123) & 0.230 (0.175) & 0.260 (0.125) & 0.226 (0.119) & 0.248 (0.129) & 0.224 (0.119) \\
\cline{2-8}
\textit{SRS} & \multicolumn{7}{c}{\textit{0.034 (0.004)}} \\

\bottomrule
\multicolumn{8}{c}{ Vulnerability prediction } \\
\toprule
2UPS & 0.069 (0.058) & 0.079 (0.052) & 0.055 (0.006) & 0.058 (0.011) & 0.074 (0.063) & 0.051 (0.007) & 0.074 (0.051) \\
DeepEST & 0.085 (0.015) & 0.084 (0.014) & 0.089 (0.024) & 0.093 (0.019) & 0.093 (0.025) & 0.096 (0.015) & 0.087 (0.018) \\
GBS & 0.034 (0.006) & 0.037 (0.006) & 0.037 (0.004) & 0.036 (0.006) & 0.035 (0.005) & \cellcolor{gray!40} \better{0.032 (0.004)} & 0.036 (0.003) \\
RHCS & 0.125 (0.016) & 0.148 (0.031) & 0.167 (0.030) & 0.160 (0.023) & 0.139 (0.017) & 0.165 (0.038) & 0.146 (0.026) \\
SSRS & 0.041 (0.017) & 0.059 (0.023) & \better{0.029 (0.004)} & \cellcolor{gray!20} \better{0.032 (0.002)} & 0.043 (0.015) & 0.034 (0.004) & 0.050 (0.023) \\
SUPS & 0.037 (0.006) & \cellcolor{gray!40} \better{0.031 (0.007)} & 0.034 (0.004) & 0.037 (0.005) & 0.036 (0.011) & 0.039 (0.007) & 0.037 (0.008) \\
\textbf{Balanced} & 0.089 (0.023) & 0.067 (0.029) & 0.057 (0.009) & 0.052 (0.010) & 0.089 (0.023) & 0.051 (0.008) & 0.061 (0.028) \\
\textbf{D2Strat} & \cellcolor{gray!20} \better{0.032 (0.003)} & \cellcolor{gray!40} \better{0.031 (0.005)} & 0.034 (0.003) & \cellcolor{gray!20} \better{0.032 (0.002)} & \cellcolor{gray!40} \better{0.031 (0.006)} & 0.034 (0.002) & \cellcolor{gray!20} \better{0.032 (0.005)} \\
\textbf{KCenterHT} & \cellcolor{gray!40} \better{0.031 (0.003)} & \cellcolor{gray!20} \better{0.032 (0.003)} & \cellcolor{gray!40} \better{0.032 (0.003)} & \cellcolor{gray!40} \better{0.030 (0.005)} & \cellcolor{gray!20} \better{0.032 (0.005)} & \cellcolor{gray!20} \better{0.033 (0.004)} & \cellcolor{gray!40} \better{0.028 (0.003)} \\
\textbf{MCUCB} & 0.042 (0.011) & 0.042 (0.011) & 0.041 (0.004) & 0.043 (0.008) & 0.042 (0.011) & 0.042 (0.005) & 0.042 (0.011) \\
\textbf{PPSsys} & 0.075 (0.073) & 0.083 (0.096) & 0.115 (0.105) & 0.061 (0.010) & 0.075 (0.073) & 0.077 (0.027) & 0.053 (0.042) \\
\textbf{Pivotal} & 0.284 (0.103) & 0.281 (0.115) & 0.327 (0.051) & 0.282 (0.077) & 0.284 (0.104) & 0.159 (0.020) & 0.288 (0.095) \\
\cline{2-8}
\textit{SRS} & \multicolumn{7}{c}{\textit{0.034 (0.002)}} \\
\bottomrule
\end{tabular}
}
\end{table}

\section{Results}
\label{sec:results}

\subsection{RQ1: Accuracy estimation error}

Table~\ref{tab:rq1-clone-RMSE} reports accuracy estimation error as the median RMSE across all models and R=30 independent runs for each strategy–feature combination (with variability shown in parentheses).
Lower RMSE indicates more accurate estimation.
For reference, we also report \textit{SRS} (simple random sampling), which does not use predictive features and serves as baseline to prior TCS studies.

\textbf{Code clone detection.}
For clone detection, the most accurate estimators are the diversity-oriented
stratified strategies \texttt{KCenterHT} and \texttt{D$^2$Strat}.
Both strategies consistently achieve the lowest RMSE across predictive features,
with \texttt{KCenterHT} yielding RMSE around $0.02$ under all features and
\texttt{D$^2$Strat} showing similarly stable behavior.
\emph{Both strategies consistently outperform \texttt{SRS} (RMSE=0.025), indicating that balanced, risk-aware stratification improves accuracy estimation beyond random sampling.}
Several classical strategies are competitive only under specific features
(e.g., \texttt{GBS} and \texttt{SSRS} with \emph{DSA}/\emph{LSA}),
whereas \texttt{RHCS} and \texttt{Pivotal} incur substantially larger errors,
indicating poor estimation accuracy.
Across strategies, representation-based features (\emph{DSA}, \emph{LSA},
\emph{MDSA}) most frequently support the lowest RMSE.

\textbf{Technical debt detection.}
The strongest accuracy estimation again comes from
\texttt{KCenterHT} and \texttt{D2Strat}.
Across features, \texttt{KCenterHT} attains RMSE close to $0.021$--$0.030$, with
its best value at \emph{DSA} ($0.021$) and similarly low values for
\emph{LSA}/\emph{MDSA} ($0.025$) and \emph{Confidence}/\emph{Entropy} ($0.028$--$0.029$).
\texttt{D2Strat} is comparably strong, reaching $0.022$ under \emph{DSA} and
$0.024$ under \emph{LSA}.
\emph{In contrast, several classical TCS strategies do not consistently outperform \textit{SRS} (RMSE $0.034$), and in some cases yield higher estimation error.}
Among classic approaches, \texttt{GBS} is consistently competitive ($0.031$--$0.035$),
and \texttt{SUPS} is also stable around $0.031$--$0.035$.
By contrast, \texttt{Pivotal} has much higher RMSE (about $0.22$--$0.26$),
showing poor accuracy estimation in this task.

\textbf{Vulnerability prediction.}
The lowest RMSE values are again achieved by
\texttt{KCenterHT} and \texttt{D2Strat} (Table~\ref{tab:rq1-clone-RMSE}).
\texttt{KCenterHT} attains the best result in this task under \emph{Entropy}
(RMSE $0.028$) and remains low under \emph{LSA} ($0.030$) and
\emph{Confidence} ($0.031$).
\texttt{D2Strat} is consistently close, with RMSE around $0.031$--$0.034$ across
features (e.g., $0.031$ under \emph{DeepGini} and \emph{Margin}, and $0.032$ under
\emph{LSA} and \emph{Entropy}).
\emph{Both strategies outperform simple random sampling (\textit{SRS}, RMSE $0.034$), but the margin of improvement is smaller than in the other tasks.}
Several classical strategies are also competitive in this task.
For example, \texttt{SUPS} reaches $0.031$ under \emph{DeepGini}, \texttt{GBS}
reaches $0.032$ under \emph{MDSA}, and \texttt{SSRS} reaches $0.029$ under
\emph{DSA}.
In contrast, \texttt{RHCS} and \texttt{Pivotal} have substantially higher errors
(above $0.12$ and $0.28$, respectively).
Across predictive features, the smallest errors in this task most often appear
under representation-oriented signals (\emph{Entropy}, \emph{LSA}, \emph{MDSA}),
with \emph{DeepGini} and \emph{DSA} also yielding strong results for specific
strategy–feature combinations.

\textbf{Cross-task patterns.}
Across all three tasks, two consistent patterns emerge.
First, balanced stratified strategies that explicitly promote diversity
(\texttt{KCenterHT} and \texttt{D$^2$Strat}) are repeatedly among the most accurate
estimators and exhibit the lowest variability across tasks. Second, predictive features aligned with representation and coverage tend to support the lowest estimation error.
Across tasks, \emph{DSA} and \emph{LSA} most frequently correspond to the smallest
RMSE, with \emph{MDSA} also competitive in clone and technical debt detection.
In vulnerability prediction, \emph{Entropy} becomes more prominent and yields
the best or near-best estimates for several strategies (notably
\texttt{KCenterHT}). At the same time, the relative importance of individual features varies by task,
indicating that prior findings from vision-based benchmarks do not transfer
uniformly to all code-related tasks.

\textbf{Illustrative examples.}
Representative per-model results (not shown in Table~\ref{tab:rq1-clone-RMSE} due to page limits) further support these patterns.
On BigCloneBench, \texttt{KCenterHT} guided by \emph{Confidence} achieves very low per-model RMSE (e.g., $0.005$ on \texttt{ModernBERT}).
On Tesoro, \texttt{KCenterHT} with \emph{Confidence} yields stable per-model errors (e.g., RMSE around $0.02$ on \texttt{GraphCodeBERT}), consistent with its low task-level RMSE in Table~\ref{tab:rq1-clone-RMSE}.
For Devign, combinations such as \texttt{KCenterHT} with \emph{LSA} also produce low per-model error (e.g., RMSE around $0.02$–$0.03$ on \texttt{StarEncoder}), aligning with the low RMSE values observed for \emph{LSA}/\emph{Entropy} in vulnerability prediction.
These examples illustrate that the aggregated trends reported in the tables are also reflected at the individual model level; complete per-model results are provided in the replication package.

\begin{figure}[t]
    \centering
    \begin{subfigure}[t]{0.48\linewidth}
        \centering
        \includegraphics[width=0.75\linewidth]{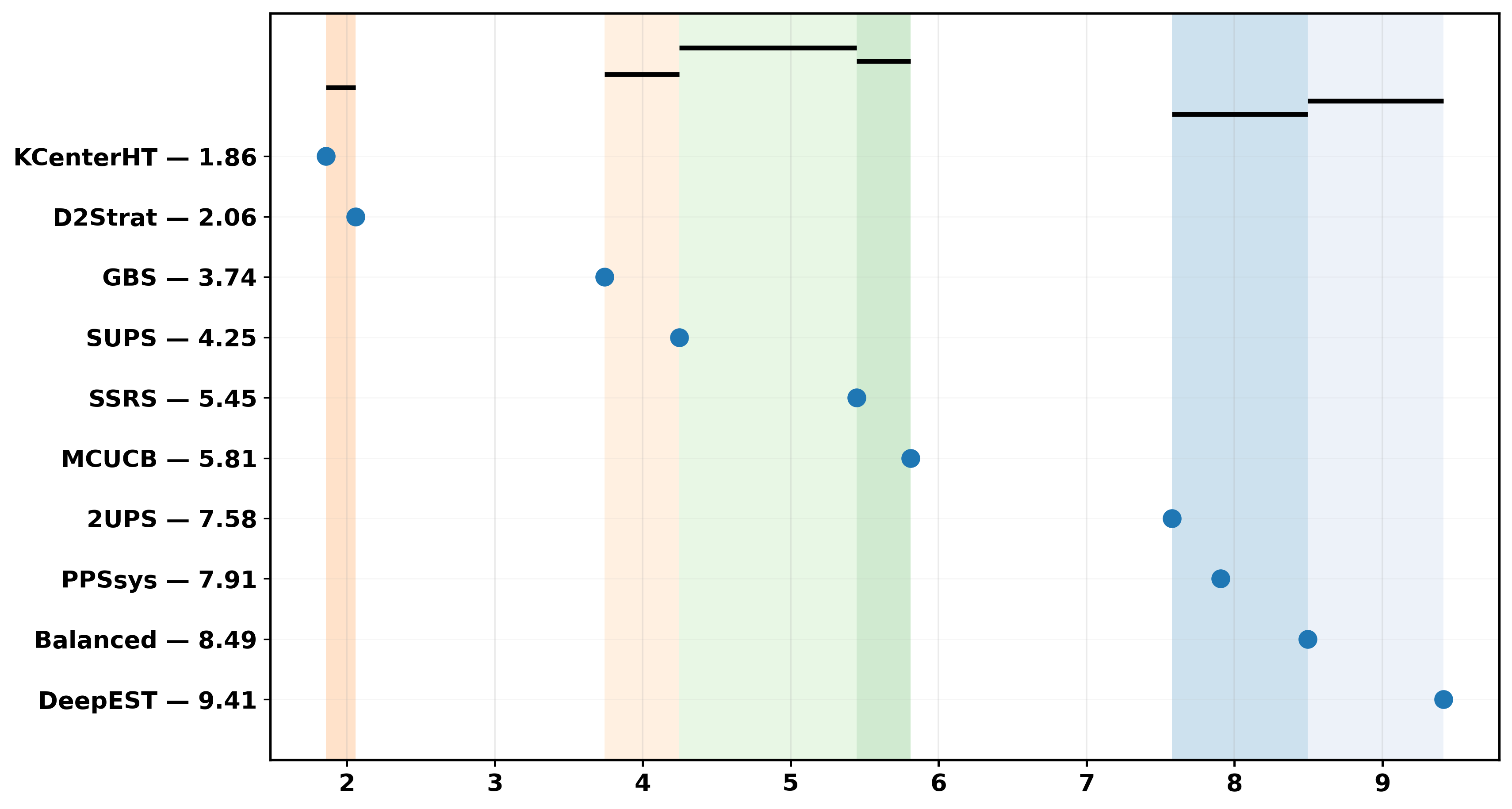}
        \caption{Strategies (RMSE).}
        \label{fig:global_rankgroups_strategy_RMSE}
    \end{subfigure}
    \hfill
    \begin{subfigure}[t]{0.48\linewidth}
        \centering
        \includegraphics[width=\linewidth]{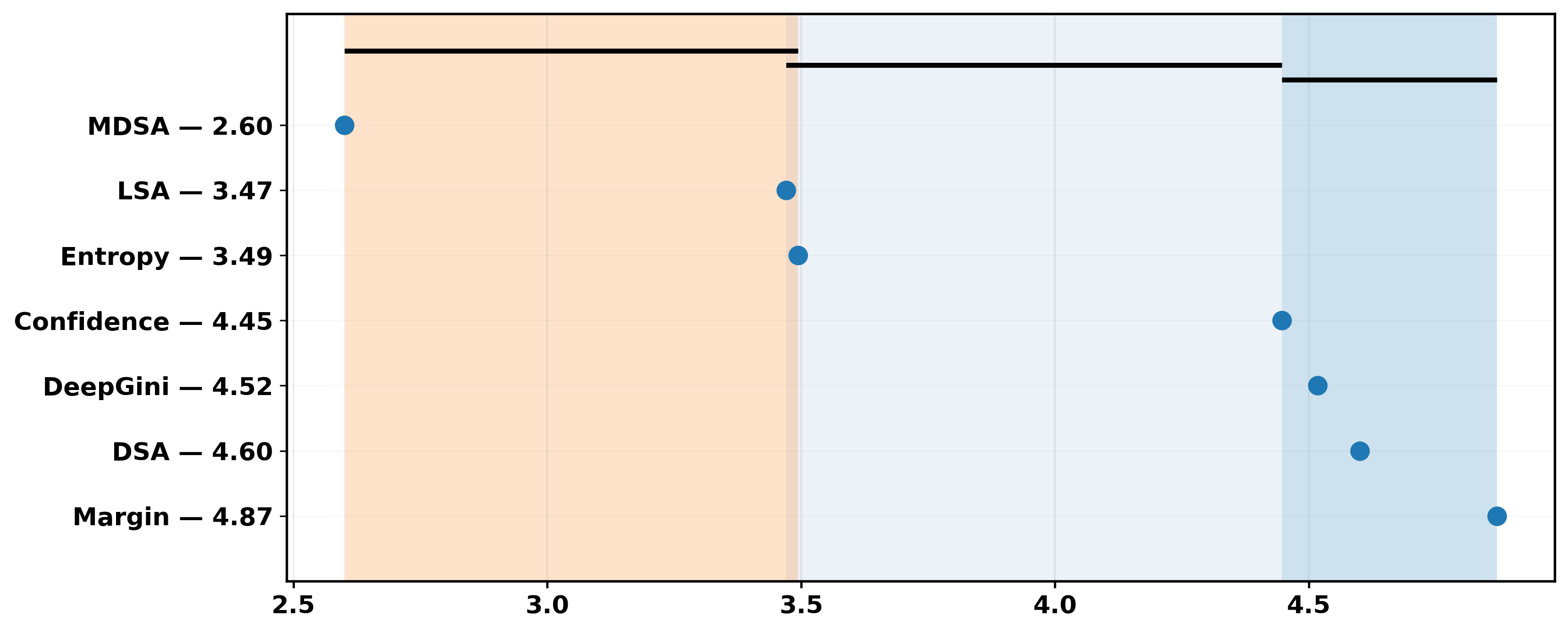}
        \caption{Predictive features (RMSE).}
        \label{fig:global_rankgroups_aux_RMSE}
    \end{subfigure}
    \caption{Friedman + post-hoc Nemenyi comparisons for error estimation (Top 10).}
    \label{fig:global_rankgroups_RMSE_pair}
\end{figure}


\textbf{Statistical analysis} The Friedman test reveals statistically significant differences among test case selection strategies with respect to accuracy estimation error when aggregating results across all models, tasks, and evaluation budgets ($p$-value$<0.001$). An equally significant effect is observed for predictive features ($p$-value$<0.001$).
Figures~\ref{fig:global_rankgroups_strategy_RMSE} and~\ref{fig:global_rankgroups_aux_RMSE} report the average ranks produced by the Friedman test together with the post-hoc Nemenyi analysis.
Lower average ranks correspond to lower accuracy estimation error. Strategies or features whose rank differences are smaller than the critical distance determined by the Nemenyi test are not statistically distinguishable and are therefore shown within the same shaded background band/area.

According to Figure~\ref{fig:global_rankgroups_strategy_RMSE}, \texttt{KCenterHT} and \texttt{D$^2$Strat} achieve the lowest average ranks and form the leading significance group. Their performance differences are not statistically significant with respect to each other, but both are significantly better than all other alternative strategies. This result is consistent with the per-task analysis at the standard budget $n{=}200$, where the same strategies repeatedly achieve the lowest RMSE. \texttt{GBS} and \texttt{SSRS} occupy intermediate ranks and are not consistently distinguishable from the top group, whereas \texttt{RHCS} and \texttt{Pivotal} are consistently ranked last, with significantly higher estimation error.
Figure~\ref{fig:global_rankgroups_aux_RMSE} shows the corresponding ranking of predictive features. Representation-based features—most notably \emph{LSA} and \emph{MDSA}—achieve the best average ranks and form the top significance group. \emph{Entropy} performs competitively, particularly for vulnerability prediction, but is statistically worse than \emph{LSA}/\emph{MDSA} in the aggregated analysis. While \emph{DSA} performs well in task-specific settings (e.g., at $n{=}200$), its lower global rank indicates that this advantage does not persist uniformly across all budgets and models.
Uncertainty-based features such as \emph{Margin} and  \emph{Confidence} rank significantly lower than the top features.

\begin{tcolorbox}[
  colback=gray!8,
  colframe=black,
  title=\textbf{Finding RQ1: Accuracy Estimation},
  boxsep=1pt,
  left=1pt,
  right=1pt,
  top=0pt,
  bottom=0pt,
  before skip=5pt,
  after skip=0pt
]
Findings from DNN testing only partially replicate for LLMs for code.
Several previously proposed TCS strategies do \emph{not} consistently outperform
simple random sampling (SRS) in terms of accuracy estimation.
In contrast, statistical sampling strategies
(\texttt{KCenterHT}, \texttt{D$^2$Strat}) reliably achieve lower and more stable
estimation error than both SRS and other TCS techniques.
Representation-based features (\emph{LSA}, \emph{MDSA}) are generally more effective
than uncertainty-only measures, though their relative benefit is task-dependent.
\end{tcolorbox}

\subsection{RQ2: Failure detection effectiveness}

Table~\ref{tab:rq2-clone-Failure} reports failure discovery at budget $n{=}200$,
shown as the \emph{median} number of detected failures (with inter-quartile range in parentheses)
computed across all models and $R{=}30$ independent runs for each strategy--feature combination.
Higher values indicate stronger failure discovery.
For reference, we also report \textit{SRS} (simple random sampling), which does not use predictive features and serves as the baseline in prior TCS studies.

\begin{table}[t]
\centering
\caption{\textcolor{black}{RQ2: Failure discovery (median/IQR) for all three tasks at $n=200$. Darker/lighter shading marks the best/second-best strategy for each feature; bold strategy names are new TCS techniques, and bold values outperform the \textit{SRS} baseline.}}
\label{tab:rq2-clone-Failure}
\scriptsize
\resizebox{.85\textwidth}{!}{
\begin{tabular}{lccccccc}
\bottomrule
\multicolumn{8}{c}{ Code Clone} \\
\toprule
Strategy & Confidence & DeepGini & DSA & LSA & Margin & MDSA & Entropy \\
\toprule

2UPS & \better{51.70 (39.92)} & \better{49.68 (39.16)} & \better{56.83 (48.93)} & \better{44.11 (36.40)} & \better{51.00 (39.23)} & \better{46.156 (42.21)} & \better{48.27 (38.26)} \\
\textbf{\textcolor{black}{Balanced}} & \better{60.20 (26.48)} & \better{57.43 (31.37)} & \better{58.37 (51.10)} & \better{45.30 (36.99)} & \better{60.20 (26.48)} & \better{47.34 (43.73)} & \better{51.02 (33.55)} \\
\textbf{\textcolor{black}{D2Strat}} & \better{42.14 (33.03)} & 40.43 (32.46) & 39.18 (33.41) & 39.55 (34.09) & \better{40.96 (31.52)} & 39.67 (34.39) & \better{40.91 (32.72)} \\
DeepEST & \cellcolor{gray!20} \cellcolor{gray!20}\better{76.25 (18.31)} & \cellcolor{gray!20} \better{71.34 (23.07)} & \cellcolor{gray!20} \better{88.40 (75.61)} & \cellcolor{gray!20} \better{63.02 (56.27)} & \cellcolor{gray!20} \better{76.55 (17.83)} & \better{72.65 (62.13)} & \cellcolor{gray!20} \better{69.16 (28.91)} \\
GBS & \better{42.34 (35.78)} & \better{41.11 (34.42)} & \better{46.64 (39.97)} & \better{41.70 (34.79)} & \better{42.55 (35.96)} & \better{42.28 (36.15)} & \better{41.28 (34.19)} \\
\textbf{\textcolor{black}{KCenterHT}} & \better{41.73 (31.22)} & \better{40.88 (31.92)} & 39.78 (34.75) & 39.01 (33.24) & \better{42.08 (31.86)} & 39.87 (34.19) & \better{41.30 (32.87)} \\
\textbf{\textcolor{black}{MCUCB}} & \better{45.17 (37.46)} & \better{45.81 (37.83)} & \better{62.03 (57.06)} & \better{46.94 (38.92)} & \better{45.86 (37.89)} & \better{50.63 (44.81)} & \better{45.12 (37.40)} \\
\textbf{\textcolor{black}{PPSsys}} & \better{59.65 (29.08)} & \better{57.52 (28.96)} & \better{57.53 (50.07)} & \better{44.244 (36.73)} & \better{59.64 (29.07)} & \better{46.03 (43.48)} & \better{52.10 (33.94)} \\
\textbf{\textcolor{black}{Pivotal}} & \better{41.76 (35.36)} & \better{42.77 (35.94)} & 22.03 (18.43) & 35.21 (33.41) & \better{41.85 (35.44)} & 35.58 (30.45) & \better{42.37 (36.06)} \\
RHCS & \cellcolor{gray!40} \better{83.38 (4.13)} & \cellcolor{gray!40} \better{84.53 (4.76)} & \cellcolor{gray!40} \better{96.75 (78.45)} & \cellcolor{gray!40} \better{70.80 (66.58)} & \cellcolor{gray!40} \better{91.52 (8.13)} & \cellcolor{gray!40} \better{84.54 (55.18)} & \cellcolor{gray!40} \better{91.92 (7.54)} \\
SSRS & \better{51.00 (33.32)} & \better{46.23 (30.41)} & 39.62 (34.74) & 40.25 (34.97) & \better{51.08 (34.04)} & 40.11 (34.87) & \better{43.78 (28.09)} \\
SUPS & \better{44.48 (34.63)} & \better{43.87 (34.50)} & \better{47.74 (40.79)} & \better{41.94 (35.15)} & \better{44.92 (35.20)} & \better{41.12 (36.47)} & \better{42.77 (34.53)} \\
\cline{2-8}
\textit{SRS} & \multicolumn{7}{c}{\textit{40.61 (34.47)}} \\
\bottomrule
\multicolumn{8}{c}{ Technical Debt} \\
\toprule

2UPS &  \cellcolor{gray!5} \better{100.07 (17.57)} & \better{98.70 (18.59)} & \better{114.65 (45.39)} & \better{99.64 (41.46)} & \cellcolor{gray!5} \better{100.47 (17.36)} & \better{98.99 (38.27)} & \cellcolor{gray!40} \better{97.67 (17.72)} \\
\textbf{\textcolor{black}{Balanced}} & \better{100.48 (18.56)} & \better{98.58 (17.71)} & \better{114.38 (45.50)} & \better{99.81 (40.45)} & \better{100.46 (18.11)} & \better{100.15 (39.13)} & \better{96.77 (18.06)} \\
\textbf{\textcolor{black}{D2Strat}} & 79.63 (25.78) & \better{79.92 (25.84)} & \better{79.84 (25.78)} & \better{79.970 (26.31)} & \better{80.66 (25.45)} & 79.05 (25.97) & 79.57 (24.97) \\
DeepEST & \cellcolor{gray!20} \better{101.80 (17.97)} & \cellcolor{gray!40} \better{102.51 (16.01)} & \cellcolor{gray!40} \better{119.34 (44.68)} & \better{105.63 (39.84)} & \cellcolor{gray!40} \better{101.90 (19.47)} & \cellcolor{gray!20} \better{108.37 (38.29)} & \better{83.72 (25.85)} \\
GBS & \better{82.60 (22.66)} & \better{83.27 (21.90)} & \better{82.96 (22.04)} & \better{92.50 (27.70)} & \better{83.10 (23.03)} & \better{88.13 (25.41)} & \better{82.56 (22.69)} \\
\textbf{\textcolor{black}{KCenterHT}} & \better{80.17 (25.55)} & \better{80.01 (25.32)} & 79.73 (25.52) & \better{80.31 (25.76)} & \better{79.87 (25.30)} & 79.00 (25.54) & \better{80.06 (25.46)} \\
\textbf{\textcolor{black}{MCUCB}} & \better{90.60 (26.74)} & \better{90.71 (26.47)} & \cellcolor{gray!45} \better{126.21 (53.96)} & \cellcolor{gray!40} \better{113.296 (39.43)} & \better{90.99 (26.49)} & \cellcolor{gray!10} \better{106.28 (40.37)} & \better{91.35 (26.37)} \\
\textbf{\textcolor{black}{PPSsys}} & \better{100.55 (17.34)} & \better{98.40 (18.25)} & \better{114.68 (45.02)} & \better{99.34 (40.00)} & \better{100.45 (17.17)} & \better{99.38 (38.01)} & \cellcolor{gray!20} \better{96.89 (18.06)} \\
\textbf{\textcolor{black}{Pivotal}} & 70.87 (29.55) & 70.67 (29.79) & 62.07 (26.52) & 67.15 (25.45) & 70.28 (29.41) & 66.80 (27.05) & 69.85 (29.66) \\
RHCS & \cellcolor{gray!40} \better{101.88 (18.06)} & \cellcolor{gray!20} \better{102.43 (16.13)} & \cellcolor{gray!20} \better{119.03 (44.82)} & \cellcolor{gray!20} \better{105.64 (39.59)} & \cellcolor{gray!20} \better{101.75 (19.76)} & \cellcolor{gray!40} \better{108.47 (38.40)} & \better{83.63 (25.99)} \\
SSRS & \better{91.72 (18.26)} & \better{85.19 (23.02)} & 67.00 (25.46) & 73.95 (25.47) & \better{91.11 (18.35)} & 78.50 (23.98) & \better{84.48 (22.58)} \\
SUPS & \better{88.83 (19.77)} & \better{88.04 (20.39)} & \better{88.69 (28.84)} & \better{85.32 (28.97)} & \better{89.14 (19.96)} & \better{86.81 (27.23)} & \better{85.92 (20.18)} \\
\cline{2-8}
\textit{SRS} & \multicolumn{7}{c}{\textit{79.77 (25.47)}} \\
\bottomrule
\multicolumn{8}{c}{ Vulnerability prediction } \\
\toprule

2UPS & \cellcolor{gray!10} \better{90.61 (11.70)} & \better{89.48 (12.34)} & \better{81.00 (16.38)} & 74.29 (18.88) &  \cellcolor{gray!10} \better{89.34 (11.46)} & 71.60 (20.23) & \better{88.11 (13.26)} \\
\textbf{\textcolor{black}{Balanced}} & \cellcolor{gray!10} \better{90.46 (10.37)} & \better{87.96 (11.56)} & \cellcolor{gray!20} \better{81.55 (17.14)} & 74.54 (19.37) & \cellcolor{gray!20} \better{90.46 (10.37)} & 72.30 (21.02) & \better{86.98 (12.54)} \\
\textbf{\textcolor{black}{D2Strat}} & 76.82 (16.47) & 76.90 (18.80) & 76.33 (17.39) & 76.85 (18.53) & 75.71 (17.58) & 76.24 (17.21) & 76.32 (18.52) \\
DeepEST & \cellcolor{gray!40} \better{94.84 (8.12)} & \cellcolor{gray!40} \better{95.24 (9.03)} & 74.87 (12.31) & 76.94 (21.15) & \cellcolor{gray!40} \better{94.52 (7.71)} & 70.60 (29.20) & \cellcolor{gray!40} \better{94.58 (9.25)} \\
GBS & \better{80.34 (15.95)} & \better{80.87 (16.50)} & \better{77.35 (18.10)} & \better{77.86 (17.49)} & \better{80.12 (16.69)} & \cellcolor{gray!20} \better{78.00 (17.18)} & \better{80.35 (16.72)} \\
\textbf{\textcolor{black}{KCenterHT}} & 75.70 (18.24) & 76.96 (18.31) & 76.91 (17.53) & \better{77.48 (17.71)} & \better{77.28 (17.86)} & \better{77.24 (17.62)} & 76.06 (17.99) \\
\textbf{\textcolor{black}{MCUCB}} & \better{82.17 (14.45)} & \better{82.17 (14.45)} & \better{78.99 (19.19)} & \cellcolor{gray!20} \better{79.08 (17.91)} & \better{82.17 (14.45)} & \better{78.72 (16.81)} & \better{82.17 (14.45)} \\
\textbf{\textcolor{black}{PPSsys}} &  \cellcolor{gray!40} \better{89.48 (9.75)} &  \cellcolor{gray!40} \better{90.34 (12.28)} & \cellcolor{gray!40} \better{81.78 (16.30)} & 74.18 (18.96) & \better{89.48 (9.75)} & 71.69 (19.45) & \better{87.53 (12.81)} \\
\textbf{\textcolor{black}{Pivotal}} & 59.68 (25.67) & 61.29 (25.84) & 72.05 (19.31) & \cellcolor{gray!40} \better{80.900 (17.95)} & 59.67 (25.66) & \cellcolor{gray!20} \better{79.88 (17.70)} & 62.97 (24.43) \\
RHCS & \cellcolor{gray!20} \better{92.08 (9.54)} & \cellcolor{gray!20} \better{92.44 (7.74)} & 68.46 (13.99) & \better{78.45 (20.79)} & \cellcolor{gray!40} \better{93.91 (9.20)} & 72.18 (30.01) & \cellcolor{gray!40} \better{94.46 (8.33)} \\
SSRS & \better{78.64 (19.18)} & 69.69 (23.81) & 76.86 (17.99) & 76.36 (17.92) & \better{78.40 (18.80)} & 75.60 (18.22) & 66.49 (26.90) \\
SUPS & \better{82.96 (14.94)} & \better{82.02 (15.42)} & \better{78.92 (17.40)} & 75.34 (18.40) & \better{83.86 (16.33)} & 75.96 (17.72) & \better{81.40 (16.43)} \\
\cline{2-8}
\textit{SRS} & \multicolumn{7}{c}{\textit{77.04 (17.40)}} \\
\bottomrule
\end{tabular}
}
\end{table}

\textbf{Code clone detection.}
The strongest failure detection is achieved by the classic
failure-oriented strategies \texttt{RHCS} and \texttt{DeepEST}.
\texttt{RHCS} consistently attains the highest counts across features, with particularly strong results under \emph{DSA}, \emph{Margin}, and \emph{Entropy} (often above $90$ failures on average).
\texttt{DeepEST} is the most competitive alternative, and performs especially well when guided by \emph{DSA}, \emph{Confidence}, and \emph{Margin}.
\emph{Both strategies substantially outperform \textit{SRS} (40.61 failures), confirming that the failure-discovery conclusions reported for vision-based DNN testing largely replicate in the clone task.}
Among the statistical sampling strategies, \texttt{Balanced} and \texttt{PPSsys} are consistently competitive (around $55$--$60$ failures under \emph{Confidence}/\emph{Margin}), but they remain below \texttt{RHCS}/\texttt{DeepEST}.
Overall, clone failure discovery benefits most from uncertainty- and boundary-proximity signals (\emph{Confidence}, \emph{Margin}) and representation-distance signals (\emph{DSA}) when paired with failure-driven strategies.

\textbf{Technical debt detection.}
In technical debt prediction, the picture changes.
The stratified diversity-oriented strategies \texttt{KCenterHT} and \texttt{D$^2$Strat} remain close to the \textit{SRS} baseline (79.77 failures), and do not provide systematic improvements in failure discovery despite their strong estimation performance in RQ1.
Instead, the highest failure counts are achieved by adaptive and failure-oriented strategies.
\texttt{MCUCB} reaches the best overall result under \emph{DSA} (126.21 failures) and also improves under \emph{LSA}.
\texttt{DeepEST} and \texttt{RHCS} also perform strongly under \emph{DSA} (around $119$ failures), while \texttt{2UPS}, \texttt{Balanced}, and \texttt{PPSsys} are consistently above the \textit{SRS} baseline under uncertainty-oriented features.
\emph{This task therefore provides a counterexample to ``one-size-fits-all'' conclusions: strategies that excel at estimation (RQ1) do not necessarily maximize early failure exposure (RQ2).}

\textbf{Vulnerability prediction.}
For vulnerability prediction, \texttt{DeepEST} achieves the strongest failure discovery overall, with high counts under \emph{Confidence}, \emph{DeepGini}, \emph{Margin}, and \emph{Entropy} (around $94$--$95$ failures on average).
\texttt{RHCS} is close behind and similarly benefits from uncertainty-driven features, particularly \emph{Margin} and \emph{Entropy}.
Among statistical sampling strategies, \texttt{PPSsys} and \texttt{Balanced} remain consistently competitive under \emph{Confidence} and \emph{Margin} (around $89$--$90$ failures), and are also strong under \emph{Entropy}.
\emph{Most strategies outperform \textit{SRS} (77.04 failures), but the gap depends strongly on the chosen feature.}
In contrast, representation-only features (\emph{LSA}, \emph{MDSA}) are generally less effective for failure discovery in this task, except in combination with adaptive allocation (\texttt{MCUCB} under \emph{LSA}).

\textbf{Cross-task patterns.}
Across tasks, two consistent patterns emerge.
First, failure discovery depends on a strong interaction between the selection strategy and the predictive feature.
Classic failure-driven strategies (\texttt{DeepEST}, \texttt{RHCS}) are the most reliable top performers, particularly when coupled with uncertainty-oriented features.
Second, the most effective features for failure discovery are \emph{Confidence} and \emph{Margin}, with \emph{DSA} providing strong task-specific gains (especially for clone and technical debt).
By contrast, the diversity-oriented stratified strategies that dominate RQ1 (\texttt{KCenterHT}, \texttt{D$^2$Strat}) do not consistently improve failure discovery over \textit{SRS}, highlighting that some prior conclusions do not transfer uniformly from vision-based settings to LLMs for code.

\textbf{Illustrative examples.}
Representative per-model results (not shown in Table~\ref{tab:rq2-clone-Failure} due to page limits) support these patterns.
For clone detection, \texttt{RHCS} with \emph{DSA} can detect over $120$ failures on individual models.
For vulnerability prediction, \texttt{DeepEST} under uncertainty-oriented features yields consistently high failure discovery (around $95$ failures on average).
For technical debt, adaptive allocation can dominate: \texttt{MCUCB} under \emph{DSA} yields the highest observed discovery in this task.
Complete per-model results are provided in the replication package.

\begin{figure}[t]
    \centering
    \begin{subfigure}[t]{0.48\linewidth}
        \centering
        \includegraphics[width=0.75\linewidth]{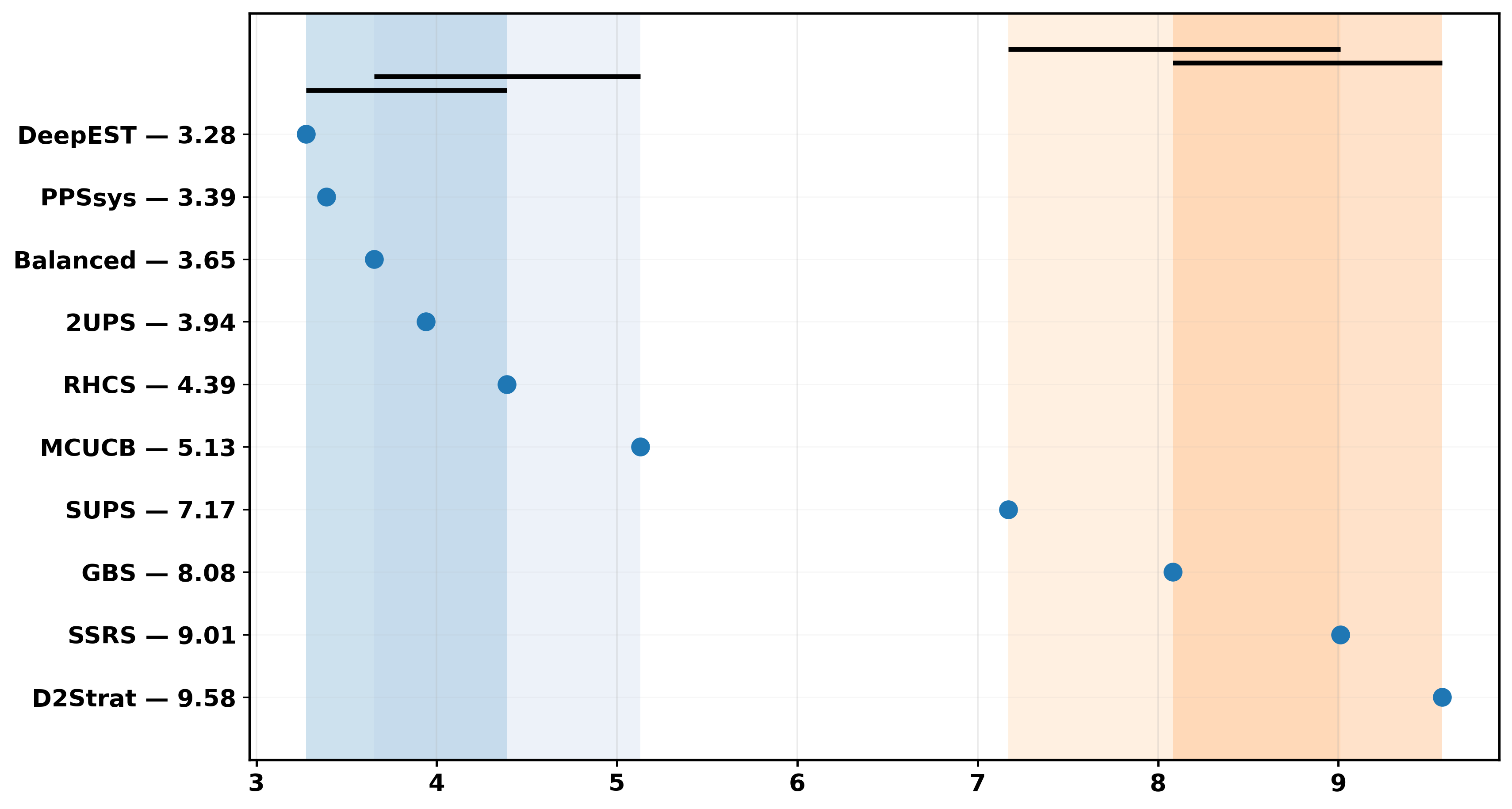}
        \caption{Test-selection strategies (failure detection).}
        \label{fig:global_rankgroups_strategy_Mean_Failures}
    \end{subfigure}
    \hfill
    \begin{subfigure}[t]{0.48\linewidth}
        \centering
        \includegraphics[width=0.96\linewidth]{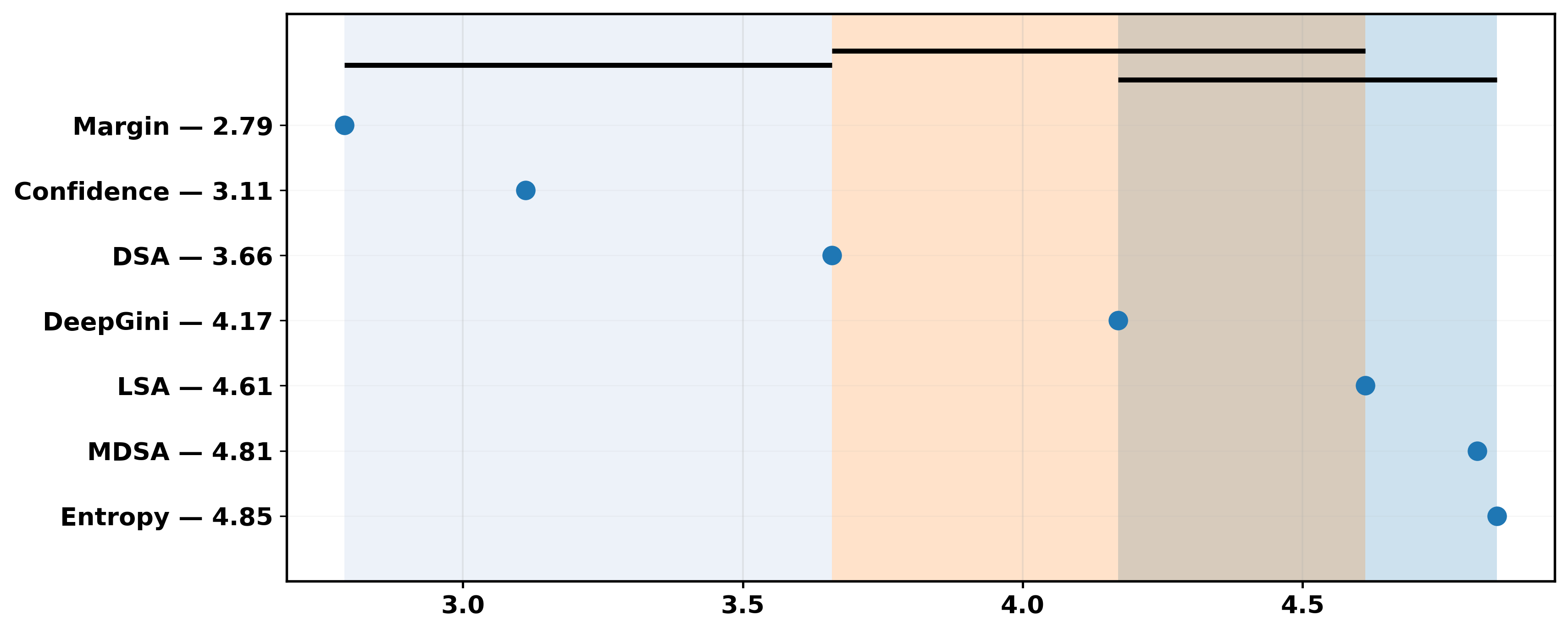}
        \caption{Predictive features (failure detection).}
        \label{fig:global_rankgroups_aux_Mean_Failures}
    \end{subfigure}
    \caption{Friedman test with post-hoc Nemenyi analysis for failure detection (Top 10).}
    \label{fig:global_rankgroups_Mean_Failures_pair}
\end{figure}

\textbf{Statistical analysis}
The Friedman test reveals statistically significant differences among test case selection strategies
with respect to failure discovery when aggregating results across all models, tasks, and evaluation budgets
($p$-value$<0.001$). A similarly significant effect is observed for predictive features ($p$-value$<0.001$).
Figures~\ref{fig:global_rankgroups_strategy_Mean_Failures} and~\ref{fig:global_rankgroups_aux_Mean_Failures}
report the average ranks produced by the Friedman test together with the post-hoc Nemenyi analysis,
where lower ranks correspond to higher failure discovery.
Strategies or features whose rank differences are smaller than the critical distance are not statistically distinguishable
and are shown within the same shaded band.

According to Figure~\ref{fig:global_rankgroups_strategy_Mean_Failures},
\texttt{DeepEST} achieves the best rank overall, with \texttt{RHCS} close behind.
\texttt{2UPS}, \texttt{Balanced}, and \texttt{PPSsys} fall in the same top significance band,
indicating that the statistical sampling strategies provide failure discovery comparable to classic failure-driven approaches.
Figure~\ref{fig:global_rankgroups_aux_Mean_Failures} shows the corresponding ranking of predictive features.
\emph{Margin} and \emph{Confidence} achieve the best average ranks and form the top significance group, with \emph{DSA} also competitive.
Representation-based features (\emph{LSA}, \emph{MDSA}) rank lower overall, although they provide task-dependent gains,
most notably when combined with adaptive allocation (e.g., \texttt{MCUCB} in technical debt prediction).
\vspace{-5pt}
\begin{tcolorbox}
[ 
colback=gray!8,
  colframe=black,
  title=\textbf{Finding RQ2: Failure Detection},
  boxsep=1pt,
  left=3pt,
  right=3pt,
  top=2pt,
  bottom=2pt,
  before skip=10pt,
  after skip=2pt
]
Failure-discovery results from vision-based DNN largely replicate for LLMs for code:
classic failure-driven strategies (\texttt{DeepEST}, \texttt{RHCS}) consistently achieve the highest discovery,
often outperforming simple random sampling (SRS).
Statistical sampling strategies (\texttt{Balanced}, \texttt{PPSsys}, \texttt{2UPS}) can reach comparable discovery under uncertainty-oriented features,
but statistical sampling methods (\texttt{KCenterHT}, \texttt{D$^2$Strat}) do not consistently improve failure discovery over SRS.
Across tasks, \emph{Confidence} and \emph{Margin} are the most reliable features, with \emph{DSA} providing strong task-specific gains, especially for technical debt detection, and~\texttt
{MCUCB}.
\end{tcolorbox}
\section{Discussion}
\label{sec:discussion}

\paragraph{\textcolor{black}{Replication and extension from vision-based DNN classification to LLM-based code classification}}
\textcolor{black}{
This work serves as a replication and extension of prior empirical evidence on
Test Case Selection (TCS), moving from DNN-based classification settings
to LLM-based code classification tasks, including vulnerability prediction, code clone detection, and
technical debt analysis.} We evaluate the same family of classical TCS strategies studied in earlier work under predictive features derived from model uncertainty and representation-space distances: \texttt{SRS} (simple random sampling),
\texttt{GBS} \cite{lv2014asymptotic,guerriero2024deepsample},
\texttt{SSRS} \cite{lohr2021sampling}, \texttt{SUPS}
\cite{lohr2021sampling,hansen1943theory}, \texttt{RHCS}
\cite{rao1962simple}, \texttt{DeepEST} \cite{guerriero2021operation}, and
\texttt{2-UPS} \cite{guerriero2024deepsample}.

DeepSample~\cite{guerriero2024deepsample} evaluates eight core TCS strategies
in DNN classification using multiple neural models and image datasets. It
reports that for accuracy estimation (RMSE, RMedSE), \texttt{SSRS} is often
among the top techniques, followed by \texttt{GBS} and frequently
\texttt{SRS}, while \texttt{DeepEST} is not competitive for estimation
accuracy. For failure exposure, \texttt{SUPS}, \texttt{DeepEST}, and
\texttt{RHCS} dominate, and for combined trade-off criteria,
\texttt{SUPS} and \texttt{RHCS} appear most often at the top. That study also
shows that predictive features strongly influence results: \emph{DSA} is most
often associated with the best estimation accuracy, while \emph{Confidence}
most often yields the highest failure counts. Adaptive probabilistic TCS for LLM sentiment analysis
\cite{asgari2025adaptive} reports similar objective-dependent behavior.
Strategies such as \texttt{GBS}, \texttt{SRS}, and \texttt{RHCS} are most
effective for error minimization with entropy-based features, while
\texttt{GBS}, \texttt{SRS}, and \texttt{2-UPS} perform best with Confidence.
For failure exposure, \texttt{SUPS}, \texttt{DeepEST}, and \texttt{SSRS}
lead, with \texttt{RHCS} strengthened under Entropy. 
\textcolor{black}{Compared to these studies, our evaluation broadens the scope by considering 17 task-specific fine-tuned model instances derived from 12 base models across three SE tasks. We include strategies replicated from prior DNN-TCS studies and six additional adaptive and
diversity-driven designs.}
This setting allows us to assess whether prior TCS
findings from DNN classification and single-model LLM studies transfer to
deep code model scenarios.

Overall, our results confirm the same core principle: strategy effectiveness is
objective-dependent, but the ranking of the strongest techniques shifts under
code-focused tasks. For estimation accuracy, the best global performance is
achieved by the diversity-oriented stratified strategies \texttt{KCenterHT} and
\texttt{D$^2$Strat} from the statistical sampling literature~\cite{arthur2006k,gonzalez1985clustering}, which consistently obtain the lowest RMSE across tasks,
budgets, and predictive features. 
\textcolor{black}{Among the strategies previously evaluated in DNN-TCS studies}, \texttt{GBS},
\texttt{SUPS}, and \texttt{SSRS} remain the most competitive for estimation,
in line with prior evidence. In contrast to DeepSample results~\cite{guerriero2024deepsample},
however, \texttt{SRS} does not appear in the top statistical group for code tasks, indicating that structured and diversity-aware sampling is
more reliable than pure random sampling for code-oriented LLM tasks. For failure discovery, our findings closely agree with prior DNN and LLM
studies. Across tasks and budgets, \texttt{DeepEST} and \texttt{RHCS} achieve
the strongest global ranks for failure detection, consistent with earlier
reports \cite{guerriero2024deepsample,asgari2025adaptive}. We also confirm
the strong role of unequal-probability and stratified designs such as
\texttt{SUPS}. Newer strategies (\texttt{Balanced},
\texttt{PPSsys}, \texttt{MCUCB}) reach comparable failure discovery levels when
paired with suitable predictive features, extending prior TCS evidence to LLM-based code classification. 

Regarding predictive features, our results replicate prior observations and extend them under a broader feature space. DeepSample compares only \emph{Confidence}, \emph{LSA}, and \emph{DSA} and reports that: for RMSE (accuracy), \emph{DSA} ranks highest, followed by \emph{LSA}, while \emph{Confidence} is weaker; for failure detection, \emph{Confidence} ranks highest, followed by \emph{DSA}, then \emph{LSA} \cite{guerriero2024deepsample}. The study on LLMs for sentiment analysis~\cite{asgari2025adaptive} evaluates only \emph{Entropy} and \emph{Confidence}, and reports that Entropy ranks stronger for error minimization, with both features effective for failure exposure. In contrast, under our broader feature set (adding \emph{Entropy}, \emph{Margin}, \emph{DeepGini}, and \emph{MDSA}), the ranking shifts. For failure detection, uncertainty features remain dominant — \emph{Confidence}, \emph{Margin}, and \emph{Entropy} are consistently top-ranked — confirming prior trends. \textcolor{black}{
However, for RMSE, our global ranking differs from DeepSample: \emph{LSA}, \emph{MDSA}, and \emph{Entropy} are most robust overall, while \emph{DSA} is strongest mainly at specific budget levels rather than consistently across budgets. This difference may be explained by the structural and semantic nature of code. Code snippets may differ syntactically while preserving the same functionality, whereas small structural changes can introduce critical errors. Representation-based features, such as LSA and MDSA, compare inputs against the model's learned representation space and can capture when a sample is unusual relative to the training distribution. They therefore perform more stably for accuracy estimation, especially when failures are linked to semantic or structural deviations rather than to prediction confidence alone.
} This side-by-side comparison shows that uncertainty features still lead failure discovery, but representation-coverage and information-based features play a larger role in accuracy estimation for code models than reported in prior DNN and single-model studies.

\textbf{Budget sensitivity}.
All techniques were evaluated under multiple testing budgets
$n$=(50, 100, 200, 400, 800).
Due to page limits, Sections~\ref{sec:results} report detailed results only for
$n$=200, which matches the budget commonly used in prior work~\cite{guerriero2021operation,guerriero2024deepsample,asgari2025adaptive}.
Across tasks and models, increasing the budget consistently increases failure
discovery and generally improves accuracy estimation, with diminishing returns
at higher budgets.
Importantly, varying the budget does not affect the main conclusions of our
study: diversity-aware statistical sampling strategies remain the most reliable
for accuracy estimation, while failure-oriented strategies consistently
prioritize early failure discovery.
The relative ranking of strategies and predictive features is largely stable
across budgets and is reflected in the statistical analysis, which aggregates
results over all evaluated budgets.
Full results for all budgets are available in our replication package.
\textcolor{black}{Because dataset sizes differ substantially, the same absolute budget corresponds to different sampling fractions across tasks. We interpret budget trends mainly within each task, especially at high budgets for smaller datasets, and use cross-task aggregation only to identify broad patterns.}

\begin{table}[t]
\centering
\caption{\textcolor{black}{Median runtime breakdown per 1{,}000 evaluated records across three software analytics tasks. We report the median runtime values in seconds together with the interquartile range (IQR).}}
\label{tab:runtime_breakdown_1k}
\scriptsize
\setlength{\tabcolsep}{3.5pt}
\renewcommand{\arraystretch}{0.95}

\begin{tabular}{l r r r r r r}
\toprule
\textbf{\textcolor{black}{Task}} &
\textbf{\textcolor{black}{\#Models}} &
\textbf{\textcolor{black}{Represent/Logit}} &
\textbf{\textcolor{black}{Uncertainty}} &
\textbf{\textcolor{black}{SA Features}} &
\textbf{\textcolor{black}{Other}} &
\textbf{\textcolor{black}{Total Runtime}} \\
\midrule
\textcolor{black}{Clone Detection} &
\textcolor{black}{3} &
\textcolor{black}{11.41 (8.65)} &
\textcolor{black}{0.0079 (0.0051)} &
\textcolor{black}{6.84 (3.69)} &
\textcolor{black}{0.24 (1.18)} &
\textcolor{black}{19.30 (6.55)} \\

\textcolor{black}{Vulnerability Detection} &
\textcolor{black}{5} &
\textcolor{black}{142.12 (16.68)} &
\textcolor{black}{0.0190 (0.0026)} &
\textcolor{black}{3.11 (0.34)} &
\textcolor{black}{0.58 (0.44)} &
\textcolor{black}{145.56 (15.75)} \\

\textcolor{black}{Technical Debt} &
\textcolor{black}{9} &
\textcolor{black}{183.22 (110.07)} &
\textcolor{black}{0.0084 (0.0069)} &
\textcolor{black}{0.27 (11.84)} &
\textcolor{black}{3.75 (6.40)} &
\textcolor{black}{203.78 (113.36)} \\

\midrule
\textcolor{black}{Overall} &
\textcolor{black}{17} &
\textcolor{black}{142.30} &
\textcolor{black}{0.0084} &
\textcolor{black}{3.11} &
\textcolor{black}{2.23} &
\textcolor{black}{145.56 (2.43 min)} \\
\bottomrule
\end{tabular}
\end{table}

\textbf{\textcolor{black}{Cost trade-offs.}}
\textcolor{black}{The main overhead of advanced TCS strategies compared to SRS (feature-agnostic technique) comes from computing predictive features, which is a one-time preprocessing step. As shown in Table~\ref{tab:runtime_breakdown_1k}, this overhead has an average of 19.30 seconds per 1K records for clone detection, 145.56 seconds (2.43min) for vulnerability detection, and 203.78 seconds (3.40min) for technical-debt detection. Across all 17 evaluated checkpoints, the average total runtime overhead is 145.56 seconds (2.43min) per 1K records. These results show that the main cost comes from inference and representation extraction, especially for larger checkpoints and SA-based features. In contrast, uncertainty features are calculations based on predicted probabilities and incur negligible cost once model outputs are available. After the features are computed, the selection step itself is lightweight. Therefore, feature-based TCS is most useful when this one-time preprocessing cost is justified by better accuracy estimation or earlier failure discovery. When expert labeling or manual inspection dominates the evaluation cost, as in vulnerability analysis, feature-based TCS can be worthwhile. However, SA feature extraction may become a substantial part of the total cost for very large training sets. In such cases, uncertainty features or SRS may remain preferable when they provide sufficient accuracy.}

\textbf{\textcolor{black}{SRS behavior under different sampling fractions.}}
\textcolor{black}{One potential concern with fixed absolute budgets is that SRS may appear to underperform simply because the same budget corresponds to very different sampling fractions across datasets. To examine this issue, we consider how sampling fraction and failure prevalence affect the variance of the SRS estimator. For SRS without replacement, the variance of the estimated failure rate $\hat{\theta}$ is
$\mathrm{Var}(\hat{\theta}) =
\frac{\theta(1-\theta)}{n}
\cdot
\frac{N-n}{N-1}$,
where $n$ is the sample size, $N$ is the dataset size, and $\theta$ is the true failure rate. The finite-population correction term, $\frac{N-n}{N-1}$, captures the effect of the sampling fraction. For $n=200$, this term is about $0.9995$ for BigCloneBench and $0.8413$ for Tesoro. Thus, the sampling-fraction effect is present and changes the variance by about $19\%$ between these datasets.}
\textcolor{black}{However, sampling fraction alone does not explain the observed behavior of SRS. The term $\theta(1-\theta)$ depends on failure prevalence and varies substantially across models and tasks. For example, in BigCloneBench, model accuracies range from $0.6565$ to $0.9900$, corresponding to $\theta(1-\theta)$ values from about $0.2255$ to $0.0099$. In Tesoro, accuracies range from $0.4191$ to $0.7482$, corresponding to values from about $0.2435$ to $0.1884$. Therefore, while different sampling fractions affect the absolute variance of SRS, failure prevalence and failure distribution also play important roles, especially when failures are rare or unevenly distributed. To reduce randomness in the comparison, we repeated all sampling methods, including SRS, 30 times under the same sample-size setting.}






\section{Threats to Validity}

\textit{Internal validity.}
Our experiments involve many strategies, predictive features, models, and
tasks, which increases the risk of implementation or configuration errors.
We mitigated this threat using a unified experimental pipeline, automated
runs, and repeated executions with fixed budgets. Still, undetected defects
or stochastic effects in sampling and model inference may influence results.

\textit{Construct validity.} We evaluate accuracy estimation using RMSE and failure discovery using
detected failures, following prior TCS studies
\cite{guerriero2024deepsample,guerriero2021operation,asgari2025adaptive}.
These metrics do not cover all testing objectives (e.g., cost or developer
effort). Predictive features such as Confidence, DSA, LSA, and Entropy are
proxies for input informativeness, and alternative signals could change the
observed rankings.
We assume that accuracy computed on the full test set is a reliable proxy for
operational performance.
In practice, real-world operational data may differ in distribution, and
labels in software engineering datasets may contain noise or ambiguity,
which could affect absolute estimates.

\textit{Conclusion validity.} We rely on non-parametric significance and post-hoc tests to draw our conclusions across
budgets, tasks, and models. While appropriate for multi-method comparison~\cite{japkowicz2011evaluating,devroey2023juge},
aggregation can hide task-specific effects. We provide detailed
per-task and per-budget results.

\textit{External validity.} Prior work evaluated fewer models and tasks (e.g., nine DNN classifiers in
DeepSample \cite{guerriero2024deepsample} and one LLM in sentiment analysis
\cite{asgari2025adaptive}). Our study extends evaluation to 17 LLMs and three
code-related tasks, improving coverage. However, results may not generalize to
all LLMs, datasets, or software engineering scenarios.

\section{Conclusions and Future Work}
This paper presented a large-scale replication study of Test Case Selection (TCS)
techniques for deep models applied to code-related tasks.
Our goal was to assess whether empirical findings reported for TCS strategies,
previously evaluated mainly on non--software engineering domains (e.g., vision-based classification), generalize to LLM-based code classification tasks such as clone detection,
vulnerability prediction, and technical debt analysis.

Our results show that this generalization is only partial.
Several strategies replicated from prior DNN-TCS studies do not consistently outperform simple random
sampling for accuracy estimation when applied to deep code models.
In contrast, statistical sampling strategies based on balanced and
diversity-aware designs achieve lower and more stable estimation error across
tasks, models, and budgets.
For failure detection, however, our results largely confirm prior findings:
failure-oriented strategies such as \texttt{DeepEST} and \texttt{RHCS} remain highly effective,
while newer adaptive strategies achieve comparable performance when paired with suitable predictive features.
Across tasks, uncertainty-oriented features support strong failure discovery,
whereas representation-based features are more closely associated with accurate estimation.
Overall, accuracy estimation and failure detection favor different classes of
strategies, confirming that TCS effectiveness is strongly objective-dependent.

An important direction for future work is extending test case selection to
\emph{generative} software engineering tasks, such as code generation~\cite{liguori23evaluates} and
automated program repair, where failure definitions, evaluation metrics, and
selection features might differ substantially.
Another promising direction concerns \emph{multi-agent and tool-augmented
systems}, where multiple code models interact within software
engineering workflows. Understanding how to apply TCS to such multi-agent settings remains an open research challenge.

\section*{Data Availability}
The replication package with raw data and code is available at: \\
\url{https://figshare.com/s/266fe72a670492e301d7}

\section*{Acknowledgments}

This work was conducted as part of the AI for Software Engineering (AI4SE) collaboration between JetBrains and Delft University of Technology. The authors gratefully acknowledge the financial support provided by JetBrains, which made this research possible.

\bibliographystyle{ACM-Reference-Format}
\bibliography{bibliography}

@article{liguori23evaluates,
  title={Who evaluates the evaluators? On automatic metrics for assessing AI-based offensive code generators},
  author={Liguori, Pietro and Improta, Cristina and Natella, Roberto and Cukic, Bojan and Cotroneo, Domenico},
  journal={Expert Systems with Applications},
  volume={225},
  pages={120073},
  year={2023},
  publisher={Elsevier}
}

@article{hou2023large,
  title={Large language models for software engineering: A systematic literature review},
  author={Hou, Xinyi and Zhao, Yanjie and Liu, Yue and Yang, Zhou and Wang, Kailong and Li, Li and Luo, Xiapu and Lo, David and Grundy, John and Wang, Haoyu},
  journal={arXiv preprint arXiv:2308.10620},
  year={2023}
}

@inproceedings{fan2023large,
  title={Large language models for software engineering: Survey and open problems},
  author={Fan, Angela and Gokkaya, Beliz and Harman, Mark and Lyubarskiy, Mitya and Sengupta, Shubho and Yoo, Shin and Zhang, Jie M},
  booktitle={2023 IEEE/ACM International Conference on Software Engineering: Future of Software Engineering (ICSE-FoSE)},
  pages={31--53},
  year={2023},
  organization={IEEE}
}

@inproceedings{pietrantuono2016adaptive,
  title={On adaptive sampling-based testing for software reliability assessment},
  author={Pietrantuono, Roberto and Russo, Stefano},
  booktitle={2016 IEEE 27th International Symposium on Software Reliability Engineering (ISSRE)},
  pages={1--11},
  year={2016},
  organization={IEEE}
}

@inproceedings{li2019boosting,
  title={{Boosting operational DNN testing efficiency through conditioning}},
  author={Li, Zenan and Ma, Xiaoxing and Xu, Chang and Cao, Chun and Xu, Jingwei and L{\"u}, Jian},
  booktitle={Proceedings of the 2019 27th ACM Joint Meeting on European Software Engineering Conference and Symposium on the Foundations of Software Engineering (ESEC/FSE)},
  publisher={ACM},
  pages={499--509},
  year={2019}
}

@book{lohr2021sampling,
  title={Sampling: design and analysis},
  author={Lohr, Sharon L.},
  year={2021},
  publisher={Chapman and Hall/CRC},
  address={New York}
}

@inproceedings{guerriero2021operation,
  title={Operation is the hardest teacher: estimating {DNN} accuracy looking for mispredictions},
  author={Guerriero, Antonio and Pietrantuono, Roberto and Russo, Stefano},
  booktitle={2021 IEEE/ACM 43rd International Conference on Software Engineering (ICSE)},
  pages={348--358},
  year={2021},
  organization={IEEE}
}

@inproceedings{guerriero2024deepsample,
  title={{DeepSample: DNN sampling-based testing for operational accuracy assessment}},
  author={Guerriero, Antonio and Pietrantuono, Roberto and Russo, Stefano},
  booktitle={Proceedings of the IEEE/ACM 46th International Conference on Software Engineering (ICSE)},
  publisher={ACM},
  pages={1--12},
  year={2024}
}

@inproceedings{kim2019guiding,
  title={Guiding deep learning system testing using surprise adequacy},
  author={Kim, Jinhan and Feldt, Robert and Yoo, Shin},
  booktitle={2019 IEEE/ACM 41st International Conference on Software Engineering (ICSE)},
  pages={1039--1049},
  year={2019},
  organization={IEEE}
}

@article{hansen1943theory,
  title={On the theory of sampling from finite populations},
  author={Hansen, Morris H and Hurwitz, William N},
  journal={The Annals of Mathematical Statistics},
  volume={14},
  number={4},
  pages={333--362},
  year={1943},
  publisher={JSTOR}
}

@article{horvitz1952generalization,
  title={A generalization of sampling without replacement from a finite universe},
  author={Horvitz, Daniel G. and Thompson, Donovan J.},
  journal={Journal of the American statistical Association},
  volume={47},
  number={260},
  pages={663--685},
  year={1952},
  publisher={Taylor \& Francis}
}

@article{lv2014asymptotic,
  title={On the asymptotic behavior of adaptive testing strategy for software reliability assessment},
  author={Lv, Junpeng and Yin, Bei-Bei and Cai, Kai-Yuan},
  journal={IEEE transactions on Software Engineering},
  volume={40},
  number={4},
  pages={396--412},
  year={2014},
  publisher={IEEE}
}

@inproceedings{zhou2024large,
  title={Large language model for vulnerability detection: Emerging results and future directions},
  author={Zhou, Xin and Zhang, Ting and Lo, David},
  booktitle={Proceedings of the 2024 ACM/IEEE 44th International Conference on Software Engineering: New Ideas and Emerging Results},
  pages={47--51},
  year={2024}
}

@inproceedings{chen2023diversevul,
  title={Diversevul: A new vulnerable source code dataset for deep learning based vulnerability detection},
  author={Chen, Yizheng and Ding, Zhoujie and Alowain, Lamya and Chen, Xinyun and Wagner, David},
  booktitle={Proceedings of the 26th International Symposium on Research in Attacks, Intrusions and Defenses},
  pages={654--668},
  year={2023}
}

@inproceedings{khajezade2024investigating,
  title={Investigating the Efficacy of Large Language Models for Code Clone Detection},
  author={Khajezade, Mohamad and Wu, Jie JW and Fard, Fatemeh Hendijani and Rodr{\'\i}guez-P{\'e}rez, Gema and Shehata, Mohamed Sami},
  booktitle={Proceedings of the 32nd IEEE/ACM International Conference on Program Comprehension},
  pages={161--165},
  year={2024}
}

@inproceedings{sonnekalb2022generalizability,
  title={Generalizability of code clone detection on codebert},
  author={Sonnekalb, Tim and Gruner, Bernd and Brust, Clemens-Alexander and M{\"a}der, Patrick},
  booktitle={Proceedings of the 37th IEEE/ACM International Conference on Automated Software Engineering},
  pages={1--3},
  year={2022}
}

@article{barr2014oracle,
  title={The oracle problem in software testing: A survey},
  author={Barr, Earl T and Harman, Mark and McMinn, Phil and Shahbaz, Muzammil and Yoo, Shin},
  journal={IEEE transactions on software engineering},
  volume={41},
  number={5},
  pages={507--525},
  year={2014},
  publisher={IEEE}
}

@article{yoo2012regression,
  title={Regression testing minimization, selection and prioritization: a survey},
  author={Yoo, Shin and Harman, Mark},
  journal={Software testing, verification and reliability},
  volume={22},
  number={2},
  pages={67--120},
  year={2012},
  publisher={Wiley Online Library}
}

@article{hu2024test,
  title={Test optimization in dnn testing: a survey},
  author={Hu, Qiang and Guo, Yuejun and Xie, Xiaofei and Cordy, Maxime and Ma, Lei and Papadakis, Mike and Le Traon, Yves},
  journal={ACM Transactions on Software Engineering and Methodology},
  volume={33},
  number={4},
  pages={1--42},
  year={2024},
  publisher={ACM New York, NY}
}

@inproceedings{guo2024stop,
  title={When to stop? towards efficient code generation in llms with excess token prevention},
  author={Guo, Lianghong and Wang, Yanlin and Shi, Ensheng and Zhong, Wanjun and Zhang, Hongyu and Chen, Jiachi and Zhang, Ruikai and Ma, Yuchi and Zheng, Zibin},
  booktitle={Proceedings of the 33rd ACM SIGSOFT International Symposium on Software Testing and Analysis},
  pages={1073--1085},
  year={2024}
}

@article{asgarimetamorphic,
  title={Metamorphic Testing of Deep Code Models: A Systematic Literature Review},
  author={Asgari, Ali and de Koning, Milan and Derakhshanfar, Pouria and Panichella, Annibale},
  journal={ACM Transactions on Software Engineering and Methodology},
  publisher={ACM New York, NY}
}

@inproceedings{astekin2025detecting,
  title={Detecting Technical Debt in Source Code Changes Using Large Language Models},
  author={Astekin, Merve and Goknil, Arda and Sen, Sagar and Tverdal, Simeon and Nguyen, Phu},
  booktitle={International Conference on Product-Focused Software Process Improvement},
  pages={334--352},
  year={2025},
  organization={Springer}
}

@article{deville1998unequal,
  title={Unequal probability sampling without replacement through a splitting method},
  author={Deville, Jean-Claude and Tille, Yves},
  journal={Biometrika},
  volume={85},
  number={1},
  pages={89--101},
  year={1998},
  publisher={Oxford University Press}
}

@article{madow1949theory,
  title={On the theory of systematic sampling, II},
  author={Madow, William G},
  journal={The Annals of Mathematical Statistics},
  volume={20},
  number={3},
  pages={333--354},
  year={1949},
  publisher={Institute of Mathematical Statistics}
}

@article{deville1998unequalSurvey,
  title={Unequal probability sampling without replacement},
  author={Deville, Jean-Claude and Tillé, Yves},
  journal={Survey Methodology},
  year={1998},
  volume={24},
  number={2},
  pages={157--168}
}

@book{tille2006sampling,
  title={Sampling Algorithms},
  author={Tillé, Yves},
  year={2006},
  publisher={Springer}
}

@incollection{hajek1971comment,
  title={Comment on “An essay on the logical foundations of survey sampling, Part One”},
  author={Hájek, Jaroslav},
  booktitle={Foundations of Statistical Inference},
  year={1971},
  publisher={Holt, Rinehart and Winston}
}

@article{carpentier2012adaptive,
  title={Adaptive stratified sampling for Monte-Carlo integration of differentiable functions},
  author={Carpentier, Alexandra and Munos, R{\'e}mi},
  journal={Advances in neural information processing systems},
  volume={25},
  year={2012}
}

@inproceedings{arthur2007proceedings,
  title={Proceedings of the eighteenth annual ACM-SIAM symposium on Discrete algorithms},
  author={Arthur, David and Vassilvitskii, Sergei},
  booktitle={Society for Industrial and Applied Mathematics},
  year={2007}
}

@techreport{arthur2006k,
  title={k-means++: The advantages of careful seeding},
  author={Arthur, David and Vassilvitskii, Sergei},
  year={2006},
  institution={Stanford}
}

@article{gonzalez1985clustering,
  title={Clustering to minimize the maximum intercluster distance},
  author={Gonzalez, Teofilo F},
  journal={Theoretical computer science},
  volume={38},
  pages={293--306},
  year={1985},
  publisher={Elsevier}
}

@inproceedings{feng2020deepgini,
  title={Deepgini: prioritizing massive tests to enhance the robustness of deep neural networks},
  author={Feng, Yang and Shi, Qingkai and Gao, Xinyu and Wan, Jun and Fang, Chunrong and Chen, Zhenyu},
  booktitle={Proceedings of the 29th ACM SIGSOFT international symposium on software testing and analysis},
  pages={177--188},
  year={2020}
}

@article{kim2023evaluating,
  title={Evaluating surprise adequacy for deep learning system testing},
  author={Kim, Jinhan and Feldt, Robert and Yoo, Shin},
  journal={ACM Transactions on Software Engineering and Methodology},
  volume={32},
  number={2},
  pages={1--29},
  year={2023},
  publisher={ACM New York, NY}
}

@inproceedings{asgari2025adaptive,
  title={Adaptive Probabilistic Operational Testing for Large Language Models Evaluation},
  author={Asgari, Ali and Guerriero, Antonio and Pietrantuono, Roberto and Russo, Stefano and others},
  booktitle={The 6th ACM/IEEE International Conference on Automation of Software Test},
  year={2025}
}

@incollection{svajlenko2021bigclonebench,
  title={Bigclonebench},
  author={Svajlenko, Jeffrey and Roy, Chanchal K},
  booktitle={Code Clone Analysis: Research, Tools, and Practices},
  pages={93--105},
  year={2021},
  publisher={Springer}
}

@article{zhou2019devign,
  title={Devign: Effective vulnerability identification by learning comprehensive program semantics via graph neural networks},
  author={Zhou, Yaqin and Liu, Shangqing and Siow, Jingkai and Du, Xiaoning and Liu, Yang},
  journal={Advances in neural information processing systems},
  volume={32},
  year={2019}
}

@misc{tesoro-code-dataset,
  author       = {Nam Hai Le and collaborators},
  title        = {Tesoro Code Dataset},
  howpublished = {\url{https://huggingface.co/datasets/NamCyan/tesoro-code}},
  note         = {Accessed: 2026-01-22},
  year         = {2024}
}

@inproceedings{gao2022adaptive,
  title={Adaptive test selection for deep neural networks},
  author={Gao, Xinyu and Feng, Yang and Yin, Yining and Liu, Zixi and Chen, Zhenyu and Xu, Baowen},
  booktitle={Proceedings of the 44th international conference on software engineering},
  pages={73--85},
  year={2022}
}

@article{ma2021test,
  title={Test selection for deep learning systems},
  author={Ma, Wei and Papadakis, Mike and Tsakmalis, Anestis and Cordy, Maxime and Traon, Yves Le},
  journal={ACM Transactions on Software Engineering and Methodology (TOSEM)},
  volume={30},
  number={2},
  pages={1--22},
  year={2021},
  publisher={ACM New York, NY, USA}
}

@article{sun2023robust,
  title={Robust test selection for deep neural networks},
  author={Sun, Weifeng and Yan, Meng and Liu, Zhongxin and Lo, David},
  journal={IEEE Transactions on Software Engineering},
  volume={49},
  number={12},
  pages={5250--5278},
  year={2023},
  publisher={IEEE}
}

@article{hu2025assessing,
  title={Assessing the Robustness of Test Selection Methods for Deep Neural Networks},
  author={Hu, Qiang and Guo, Yuejun and Xie, Xiaofei and Cordy, Maxime and Ma, Wei and Papadakis, Mike and Ma, Lei and Le Traon, Yves},
  journal={ACM Transactions on Software Engineering and Methodology},
  year={2025},
  publisher={ACM New York, NY}
}

@inproceedings{le2025impacts,
  title={On the impacts of contexts on repository-level code generation},
  author={Le Hai, Nam and Nguyen, Dung Manh and Bui, Nghi DQ},
  booktitle={Findings of the Association for Computational Linguistics: NAACL 2025},
  pages={1496--1524},
  year={2025}
}

@article{koohestani2025benchmarking,
  title={Benchmarking AI Models in Software Engineering: A Review, Search Tool, and Unified Approach for Elevating Benchmark Quality},
  author={Koohestani, Roham and de Bekker, Philippe and Ko{\c{c}}, Beg{\"u}m and Izadi, Maliheh},
  journal={IEEE Transactions on Software Engineering},
  year={2025},
  publisher={IEEE}
}

@article{ni2026learning,
  title={Learning-based models for vulnerability detection: An extensive study},
  author={Ni, Chao and Yin, Xin and Shen, Liyu and Wang, Shaohua},
  journal={Empirical Software Engineering},
  volume={31},
  number={1},
  pages={18},
  year={2026},
  publisher={Springer}
}

@article{feng2020codebert,
  title={Codebert: A pre-trained model for program-ming and natural languages},
  author={Feng, Z},
  journal={arXiv preprint arXiv:2002.08155},
  year={2020}
}

@inproceedings{wang2023codet5p,
  title={Codet5+: Open code large language models for code understanding and generation},
  author={Wang, Yue and Le, Hung and Gotmare, Akhilesh and Bui, Nghi and Li, Junnan and Hoi, Steven},
  booktitle={Proceedings of the 2023 conference on empirical methods in natural language processing},
  pages={1069--1088},
  year={2023}
}

@article{roziere2023code,
  title={Code llama: Open foundation models for code},
  author={Roziere, Baptiste and Gehring, Jonas and Gloeckle, Fabian and Sootla, Sten and Gat, Itai and Tan, Xiaoqing Ellen and Adi, Yossi and Liu, Jingyu and Sauvestre, Romain and Remez, Tal and others},
  journal={arXiv preprint arXiv:2308.12950},
  year={2023}
}

@article{zhang2023survey,
  title={A survey on large language models for software engineering},
  author={Zhang, Quanjun and Fang, Chunrong and Xie, Yang and Zhang, Yaxin and Yang, Yun and Sun, Weisong and Yu, Shengcheng and Chen, Zhenyu},
  journal={arXiv preprint arXiv:2312.15223},
  year={2023}
}

@article{rao1962simple,
  title={On a simple procedure of unequal probability sampling without replacement},
  author={Rao, J NoK and Hartley, HO and Cochran, WG},
  journal={Journal of the Royal Statistical Society Series B: Statistical Methodology},
  volume={24},
  number={2},
  pages={482--491},
  year={1962},
  publisher={Oxford University Press}
}

@book{japkowicz2011evaluating,
  title={Evaluating Learning Algorithms: A Classification Perspective},
  author={Japkowicz, N. and Shah, M.},
  isbn={9781139494144},
  url={https://books.google.com/books?id=VoWIIOKVzR4C},
  year={2011},
  publisher={Cambridge University Press}
}

@article{Garcia:2009,
 author = {Garc\'{\i}a, Salvador and Molina, Daniel and Lozano, Manuel and Herrera, Francisco},
 title = {A Study on the Use of Non-parametric Tests for Analyzing the Evolutionary Algorithms' Behaviour: A Case Study on the {CEC}'2005 Special Session on Real Parameter Optimization},
 journal = {Journal of Heuristics},
 issue_date = {December  2009},
 volume = {15},
 number = {6},
 month = dec,
 year = {2009},
 issn = {1381-1231},
 pages = {617--644},
 numpages = {28},
 acmid = {1666087},
}

@article{devroey2023juge,
  title={JUGE: An infrastructure for benchmarking Java unit test generators},
  author={Devroey, Xavier and Gambi, Alessio and Galeotti, Juan Pablo and Just, Ren{\'e} and Kifetew, Fitsum and Panichella, Annibale and Panichella, Sebastiano},
  journal={Software Testing, Verification and Reliability},
  volume={33},
  number={3},
  pages={e1838},
  year={2023},
  publisher={Wiley Online Library}
}

@inproceedings{mondal2015exploring,
  title={Exploring test suite diversification and code coverage in multi-objective test case selection},
  author={Mondal, Debajyoti and Hemmati, Hadi and Durocher, Stephane},
  booktitle={2015 IEEE 8th International Conference on Software Testing, Verification and Validation (ICST)},
  pages={1--10},
  year={2015},
  organization={IEEE}
}

@inproceedings{noor2015similarity,
  title={A similarity-based approach for test case prioritization using historical failure data},
  author={Noor, Tanzeem Bin and Hemmati, Hadi},
  booktitle={2015 IEEE 26th International Symposium on Software Reliability Engineering (ISSRE)},
  pages={58--68},
  year={2015},
  organization={IEEE}
}

@article{guo2020graphcodebert,
  title={Graphcodebert: Pre-training code representations with data flow},
  author={Guo, Daya and Ren, Shuo and Lu, Shuai and Feng, Zhangyin and Tang, Duyu and Liu, Shujie and Zhou, Long and Duan, Nan and Svyatkovskiy, Alexey and Fu, Shengyu and others},
  journal={arXiv preprint arXiv:2009.08366},
  year={2020}
}

@inproceedings{warner2025smarter,
  title={Smarter, better, faster, longer: A modern bidirectional encoder for fast, memory efficient, and long context finetuning and inference},
  author={Warner, Benjamin and Chaffin, Antoine and Clavi{\'e}, Benjamin and Weller, Orion and Hallstr{\"o}m, Oskar and Taghadouini, Said and Gallagher, Alexis and Biswas, Raja and Ladhak, Faisal and Aarsen, Tom and others},
  booktitle={Proceedings of the 63rd Annual Meeting of the Association for Computational Linguistics (Volume 1: Long Papers)},
  pages={2526--2547},
  year={2025}
}

@article{guo2022unixcoder,
  title={Unixcoder: Unified cross-modal pre-training for code representation},
  author={Guo, Daya and Lu, Shuai and Duan, Nan and Wang, Yanlin and Zhou, Ming and Yin, Jian},
  journal={arXiv preprint arXiv:2203.03850},
  year={2022}
}

@article{li2023starcoder,
  title={Starcoder: may the source be with you!},
  author={Li, Raymond and Allal, Loubna Ben and Zi, Yangtian and Muennighoff, Niklas and Kocetkov, Denis and Mou, Chenghao and Marone, Marc and Akiki, Christopher and Li, Jia and Chim, Jenny and others},
  journal={arXiv preprint arXiv:2305.06161},
  year={2023}
}

@article{wei2023magicoder,
  title={Magicoder: Empowering code generation with oss-instruct},
  author={Wei, Yuxiang and Wang, Zhe and Liu, Jiawei and Ding, Yifeng and Zhang, Lingming},
  journal={arXiv preprint arXiv:2312.02120},
  year={2023}
}

@inproceedings{ahmad2021unified,
  title={Unified pre-training for program understanding and generation},
  author={Ahmad, Wasi and Chakraborty, Saikat and Ray, Baishakhi and Chang, Kai-Wei},
  booktitle={Proceedings of the 2021 conference of the North American chapter of the association for computational linguistics: human language technologies},
  pages={2655--2668},
  year={2021}
}

@article{lozhkov2024starcoder,
  title={Starcoder 2 and the stack v2: The next generation},
  author={Lozhkov, Anton and Li, Raymond and Allal, Loubna Ben and Cassano, Federico and Lamy-Poirier, Joel and Tazi, Nouamane and Tang, Ao and Pykhtar, Dmytro and Liu, Jiawei and Wei, Yuxiang and others},
  journal={arXiv preprint arXiv:2402.19173},
  year={2024}
}

@article{zhang2024tinyllama,
  title={Tinyllama: An open-source small language model},
  author={Zhang, Peiyuan and Zeng, Guangtao and Wang, Tianduo and Lu, Wei},
  journal={arXiv preprint arXiv:2401.02385},
  year={2024}
}

@article{javaheripi2023phi,
  title={Phi-2: The surprising power of small language models},
  author={Javaheripi, Mojan and Bubeck, S{\'e}bastien and Abdin, Marah and Aneja, Jyoti and Bubeck, Sebastien and Mendes, Caio C{\'e}sar Teodoro and Chen, Weizhu and Del Giorno, Allie and Eldan, Ronen and Gopi, Sivakanth and others},
  journal={Microsoft Research Blog},
  volume={1},
  number={3},
  pages={3},
  year={2023}
}

@article{allamanis2018survey,
  title={A survey of machine learning for big code and naturalness},
  author={Allamanis, Miltiadis and Barr, Earl T and Devanbu, Premkumar and Sutton, Charles},
  journal={ACM Computing Surveys (CSUR)},
  volume={51},
  number={4},
  pages={1--37},
  year={2018},
  publisher={ACM New York, NY, USA}
}

@article{capobianco2013improving,
  title={Improving IR-based traceability recovery via noun-based indexing of software artifacts},
  author={Capobianco, Giovanni and Lucia, Andrea De and Oliveto, Rocco and Panichella, Annibale and Panichella, Sebastiano},
  journal={Journal of Software: Evolution and Process},
  volume={25},
  number={7},
  pages={743--762},
  year={2013},
  publisher={Wiley Online Library}
}

@inproceedings{sun2024ai,
  title={Ai coders are among us: Rethinking programming language grammar towards efficient code generation},
  author={Sun, Zhensu and Du, Xiaoning and Yang, Zhou and Li, Li and Lo, David},
  booktitle={Proceedings of the 33rd ACM SIGSOFT International Symposium on Software Testing and Analysis},
  pages={1124--1136},
  year={2024}
}

@article{panichella2021systematic,
  title={A systematic comparison of search-based approaches for LDA hyperparameter tuning},
  author={Panichella, Annibale},
  journal={Information and Software Technology},
  volume={130},
  pages={106411},
  year={2021},
  publisher={Elsevier}
}

@article{jiang2018predicting,
  title={Predicting the generalization gap in deep networks with margin distributions},
  author={Jiang, Yiding and Krishnan, Dilip and Mobahi, Hossein and Bengio, Samy},
  journal={arXiv preprint arXiv:1810.00113},
  year={2018}
}

@article{zheng2023surveycode,
  title={A survey of large language models for code: Evolution, benchmarking, and future trends},
  author={Zheng, Zibin and Ning, Kaiwen and Wang, Yanlin and Zhang, Jingwen and Zheng, Dewu and Ye, Mingxi and Chen, Jiachi},
  journal={arXiv preprint arXiv:2311.10372},
  year={2023}
}

@inproceedings{chen2023large,
  title={Large language models meet NL2Code: A survey},
  author={Zan, Daoguang and Chen, Bei and Zhang, Fengji and Lu, Dianjie and Wu, Bingchao and Guan, Bei and Yongji, Wang and Lou, Jian-Guang},
  booktitle={Proceedings of the 61st Annual Meeting of the Association for Computational Linguistics (Volume 1: Long Papers)},
  pages={7443--7464},
  year={2023}
}

@article{abbasishahkoo2025metasel,
  title={Metasel: A test selection approach for fine-tuned dnn models},
  author={Abbasishahkoo, Amin and Dadkhah, Mahboubeh and Briand, Lionel and Lin, Dayi},
  journal={IEEE Transactions on Software Engineering},
  year={2025},
  publisher={IEEE}
}

@article{aghababaeyan2024deepgd,
  title={Deepgd: A multi-objective black-box test selection approach for deep neural networks},
  author={Aghababaeyan, Zohreh and Abdellatif, Manel and Dadkhah, Mahboubeh and Briand, Lionel},
  journal={ACM Transactions on Software Engineering and Methodology},
  volume={33},
  number={6},
  pages={1--29},
  year={2024},
  publisher={ACM New York, NY}
}

@article{weinstein2020margin,
  title={Margin-based regularization and selective sampling in deep neural networks},
  author={Weinstein, Berry and Fine, Shai and Hel-Or, Yacov},
  journal={arXiv preprint arXiv:2009.06011},
  year={2020}
}

@article{ducoffe2018adversarial,
  title={Adversarial active learning for deep networks: a margin based approach},
  author={Ducoffe, Melanie and Precioso, Frederic},
  journal={arXiv preprint arXiv:1802.09841},
  year={2018}
}

@article{wang2016cost,
  title={Cost-effective active learning for deep image classification},
  author={Wang, Keze and Zhang, Dongyu and Li, Ya and Zhang, Ruimao and Lin, Liang},
  journal={IEEE Transactions on Circuits and Systems for Video Technology},
  volume={27},
  number={12},
  pages={2591--2600},
  year={2016},
  publisher={IEEE}
}
\end{document}